\documentclass[a4paper,11pt]{article}

\usepackage{jcappub} 

\usepackage{epstopdf}
\usepackage{ulem}
\usepackage{amssymb}
\usepackage{natbib}
\usepackage{times}
\usepackage{graphicx,bm,amssymb}
\usepackage{pstricks}
\usepackage{psfrag}
\usepackage{gensymb}
\usepackage{appendix}

\usepackage{amsmath}
\allowdisplaybreaks

\usepackage[colorlinks=true,linkcolor=blue,citecolor=blue]{hyperref}
\newcommand{\be}{\begin{equation}}
\newcommand{\ee}{\end{equation}}
\newcommand{\bse}{\begin{subequations}}
\newcommand{\ese}{\end{subequations}}
\newcommand{\bary}{\begin{eqnarray}}
\newcommand{\eary}{\end{eqnarray}}
\newcommand{\en}{E_\nu}

\def\aj{AJ}
\def\araa{ARA\&A}
\def\apj{ApJ}
\def\apjl{ApJ}
\def\apjs{ApJS}
\def\apss{Ap\&SS}
\def\aap{A\&A}
\def\aapr{A\&A~Rev.}
\def\mnras{MNRAS}
\def\pasa{PASA}
\def\prd{Phys.~Rev.~D}
\def\nat{Nature}
\def\physrep{Phys.~Rep.}
\def\physscr{Phys.~Scr}

\pdfoutput=1 

\title{Analysis of Fermi-LAT observations, UHECRs and neutrinos from the radio galaxy Centaurus B}
\author[a,1]{N. Fraija\note{Corresponding author.}, }
\author[b]{M. Araya,}
\author[a]{A. Galv\'an-G\'amez}
\author[a]{and J. A. de Diego}

\affiliation[a]{Instituto de Astronom\' ia, Universidad Nacional Aut\'onoma de M\'exico, Circuito Exterior, C.U., A. Postal 70-264, 04510 M\'exico City, M\'exico.}

\affiliation[b]{Escuela de F\'isica \& Centro de Investigaciones Espaciales (CINESPA), Universidad de Costa Rica, San Jos\'e 2060, Costa Rica.}

\emailAdd{nifraija@astro.unam.mx}
\emailAdd{miguel.araya@ucr.ac.cr}
\emailAdd{agalvan@astro.unam.mx}
\emailAdd{jdo@astro.unam.mx}

\abstract{Centaurus B (Cen B) is one of the closest and brightest radio-loud galaxy in the southern sky. This radio galaxy, proposed as a plausible candidate for accelerating ultra-high-energy cosmic rays (UHECRs), is near the highest-energy neutrino event reported (IC35) in the High-Energy Starting Events catalog.  Pierre Auger observatory reported the highest energy comic rays during 10 years of collecting data with some of them around this source.   In this paper, the analysis of the gamma-ray spectrum and the light curve above 200 MeV is presented with nine years of cumulative Fermi-LAT data around Cen B.  Taking into consideration the multi-wavelength observations carried out about this radio galaxy,  leptonic and hadronic scenarios are introduced in order to fit the  spectral energy distribution, assuming that the gamma-ray flux is produced in a region close to the core and in the extended lobes.  Using the best-fit values found, several physics properties of this radio galaxy are derived.  Furthermore, a statistical analysis of the cosmic ray distribution around Cen B is performed, finding that this distribution is not different from the background at a level of significance of 5\%.   Considering the UHECR event associated to this source by Moskalenko et al.  \cite{2009ApJ...693.1261M} and extrapolating its luminosity to low energies, we do not find enough evidence to associate the highest-energy neutrino event (IC35) with this radio galaxy.
}

\begin{document}

\date{\today} 

\maketitle


%
\section{Introduction}
At a distance of 56 Mpc (z=0.0129),  Centaurus B (Cen B), also known as PKS 1343-601, is one of the nearest radio-loud galaxy, which makes it an affordable object for investigating  the physics of outflows and extended  lobes. This source is the fifth brightest radio galaxy (after Cygnus A, Cen A, M87 and Fomax A) in the sky.  Because of the proximity to the Galactic plane, Cen B has been poorly studied in comparison with the closest radio galaxy Centaurus A.  Preliminary studies have associated to Cen B with one of the most significant gathering of mass (the Great Attractor)  as well as a group of galaxies hidden behind the Milky Way \citep{1988ApJ...326...19L, 1999PASA...16...53K, 2000A&ARv..10..211K}.\\
%
%
The unresolved core and the extended  lobes of Cen B have been monitored in several energy bands.  In radio wavelengths, this radio galaxy has been detected at 4.8 and 8.64 GHz with the Australia Telescope Compact Array \citep{2001MNRAS.325..817J},  at 843 MHz with Molonglo Observatory Synthesis Telescope \citep{1991PASAu...9..255M}, and at 30, 44, 70, 100 and 143 GHz with Planck satellite \citep{2011A&A...536A...1P, 2011A&A...536A...7P}.   In the X-ray band, Cen B has been detected by ASCA, Suzaku and Chandra satellites.  The ASCA collaboration reported a  diffuse emission from an extended halo that matches with the location of the  lobes of Cen B  \citep{1998ApJ...499..713T}.  The Chandra satellite collected data at 0.5 - 7 keV from the inner parts of the unresolved core \citep{2005ApJS..156...13M}.  The Suzaku satellite monitored this source from 2011 July 16 to 18, providing constraining upper limits \citep{2013A&A...550A..66K}.  In the gamma-ray band, this source is continuously observed in the survey.  Katsuta et al.   \cite{2013A&A...550A..66K} presented a specific dataset  during a period of almost 43 months  (from 2008 August 4 to 2012 February 18) by the Large Area Telescope (LAT) \citep{2009ApJ...697.1071A} instrument onboard the Fermi Satellite. \\ 
\\
%
%
Pierre Auger Observatory (PAO), located at  Malargue (Argentina), reported initially the distribution of 27 ultra-high-energy cosmic ray (UHECR) events  collected over almost 4 years of operations, from 2004 January 1$^{\rm th}$ to 2007 August 31 \citep{2007Sci...318..938P, 2008APh....29..188P}.  Taking into consideration the complete analysis performed by this observatory, and the proximity of Cen B to the Galactic plane and its magnetic field, Moskalenko et al. \citep{2009ApJ...693.1261M} proposed to Cen B as a plausible candidate for accelerating UHECRs. Later, PAO reported the UHECR events  collected over 10 years of operations, from 2004 January 1$^{\rm th}$ up to 2014 March 31 \citep{2015ApJ...804...15A}.  The arrival directions of some of the highest-energy events ($E_{\rm A}\geq 5.8 10^{19} eV$) are located near the Cen B direction.\\
\\
%
%
%
The IceCube neutrino telescope, located at the South Pole, reported in the High-Energy Starting Events (HESE) catalog\footnote{http://icecube.wisc.edu/science/data/HE-nu-2010-2014}  a sample of 82 extraterrestrial neutrino events in the TeV - PeV energies \citep{2017arXiv171001191I}.   The highest-energy neutrino reported in this catalog, the IC35 event, was 2004$^{+236}_{-262}$ TeV. This event shower-type was located with a median angular error of 15.9$^\circ$  centered at RA=$208.4^\circ$ and Dec=$-55.8^\circ$ (J2000).   Using the second catalog of Active Galactic Nuclei (AGN) reported by Fermi-LAT (2LAC) \citep{2011ApJ...743..171A}, Kadler et al. \cite{2016NatPh..12..807K}  found 20 extragalactic objects within the median angular error of IC35, being Cen A and Cen  B  the nearest AGN among the most promising candidates.   However,  the quasar PKS B1424-418, one of the 20 sources, exhibited a giant and long-lasting outburst. Taking into account the temporal coincidence of the IC35 event with this outburst,   Kadler et al. modelled the giant outburst using the photo-hadronic interactions and correlated the IC35 event with the quasar PKS B1424-418.\\
\\
In this work we analyze  nine years of  Fermi-LAT data and also model the broadband spectral energy distribution (SED) of the extended  lobes and the region close to the core of the radio galaxy Cen B.  In addition, we use statistical analysis of the UHECR distribution around this source, and finally,  we study the possible association of the PeV-neutrino event (IC35) with this radio galaxy.  The paper is arranged as follows.   In Section 2, we present the Fermi-LAT  data reduction, and show the spectral and temporal analysis. In Section 3 we model the broadband SED from the extended  lobes and the region close to the core.   In Section 4, we study the chemical composition, the necessary condition for acceleration and the statistical analysis of UHECRs.  In Section 5, we show the neutrino production and finally, in Section 6, we present our conclusions. 
\section{LAT-Fermi Data Reduction and Analysis }\label{sec:2}
\subsection{Data reduction}
Fermi-LAT is a gamma-ray telescope with sensitivity in the range from 20 MeV to $\sim 300$ GeV. We analyze data collected from August 2008 to June 2017 with the latest version of the publicly available ``ScienceTools'' (v10r0p5)\footnote{http://fermi.gsfc.nasa.gov/ssc/data}, Pass 8 event-level analysis and the instrument response functions. SOURCE class events are selected with a maximum zenith angle of 90$\degree$ and the time intervals when the data quality is optimal. The events are filtered in energy above 200 MeV in a 20$\degree \times 20 \degree$ region of interest around the coordinates RA=206.7$\degree$ and  Dec=-60.41$\degree$ (J2000).  Data below this energy are not used since the point spread function (PSF) of the instrument becomes broad and the uncertainties in the effective area and background model are large.\\ 
In order to obtain the spectral parameters and the significances of the sources,  a maximum likelihood fit as indicated in \cite{1996ApJ...461..396M} is carried out with a model including known sources from the LAT Third Source Catalog (3FGL) \citep{2015ApJS..218...23A}.   The model also includes the Galactic diffuse emission and the residual background and isotropic extragalactic emission, given respectively by the files gll\_iem\_v06.fits and iso\_P8R2\_SOURCE\_V6\_v06.txt, provided by the LAT team \footnote{See https://fermi.gsfc.nasa.gov/ssc/data/access/lat/BackgroundModels.html}.   Data are binned spatially with a scale of 0.1$\degree$ per pixel and ten logarithmically spaced bins per decade in energy. The significance of each source is estimated as the square root of the test statistic (TS) defined as $-2\times$log$(L_0/L)$, with $L_0$ and $L$ the likelihood values obtained before and after including the source in the model, respectively.
\subsection{Spectral analysis}
\begin{figure}
\vspace{0.4cm}
{\centering
\resizebox*{0.49\textwidth}{0.3\textheight}
{\includegraphics{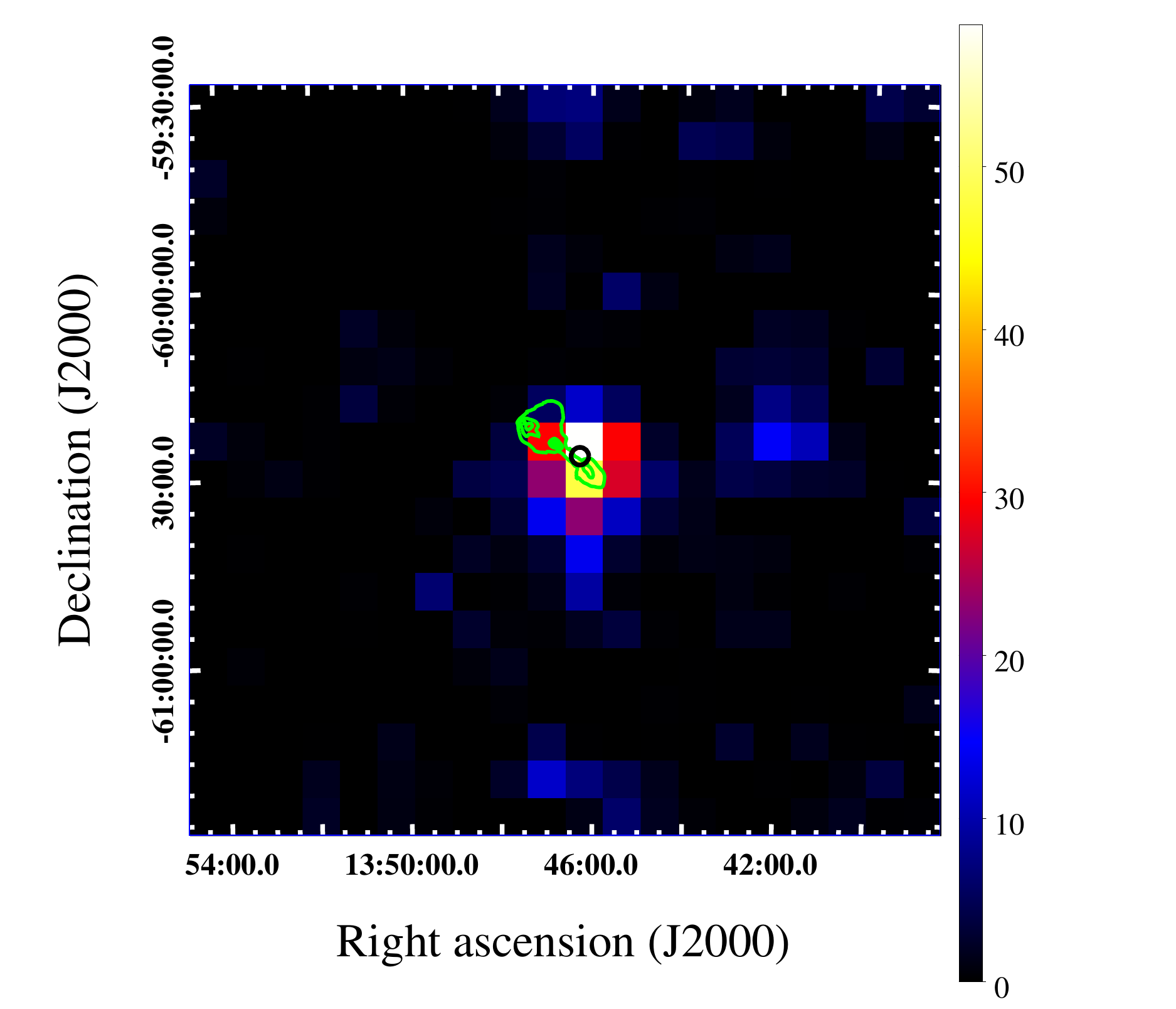}}
\resizebox*{0.49\textwidth}{0.3\textheight}
{\includegraphics{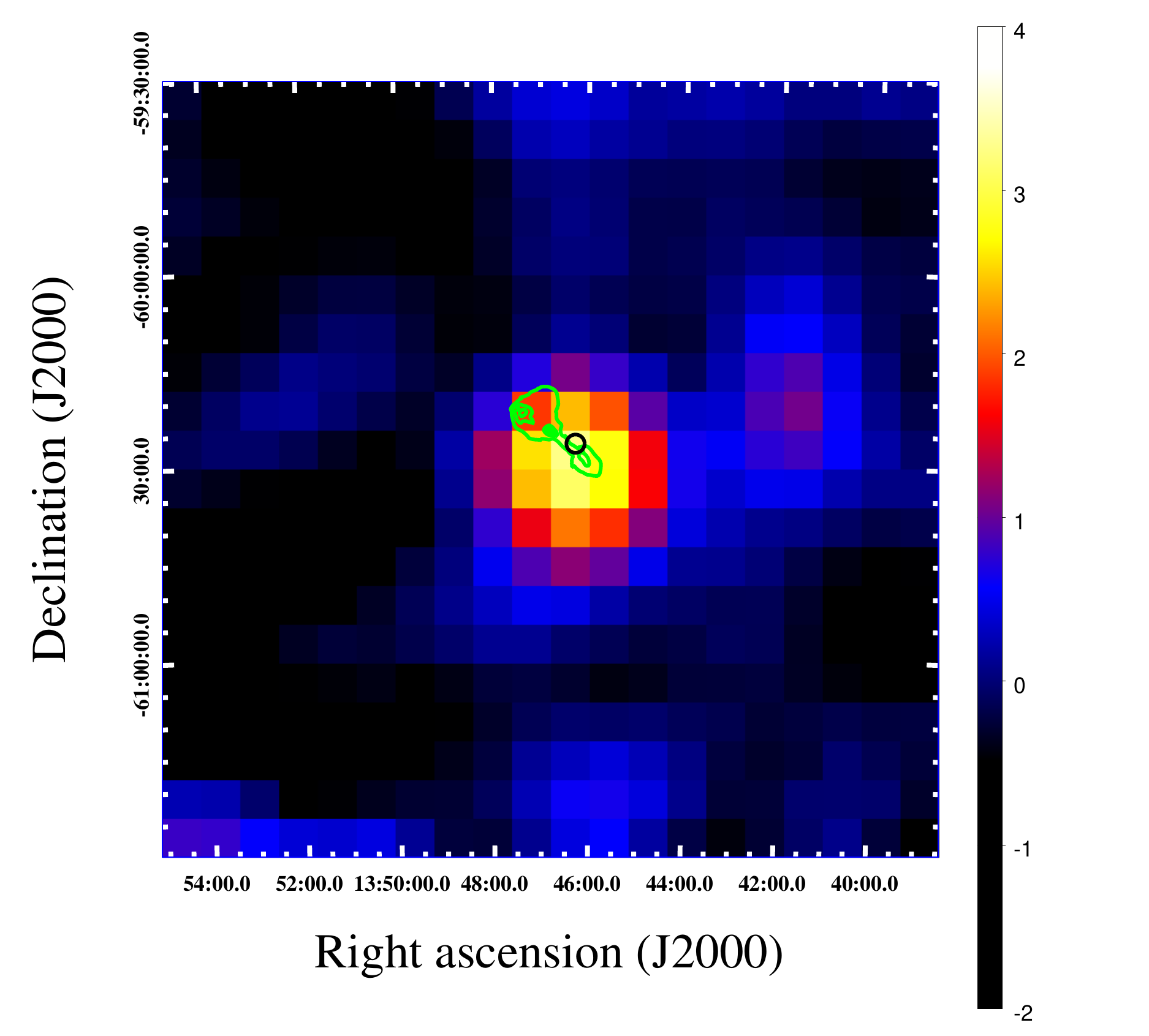}}

}
\caption{The left-hand panel: TS map for a point source hypothesis obtained above 5 GeV with the improved LAT model. The map clearly shows the excess associated to Cen B. The radio contours are taken from an 843 MHz SUMSS observation (see text) and are shown at the 0.4, 1.3, 2.2, 3.1 and 4 Jy/beam. No other 3FGL sources are found within the region shown. The black circle represents the 68\% confidence level error region for Cen B found in the analysis. The right-hand panel: Residual count map obtained with the data shown in the left-hand panel and showing the same region and contour levels. This image has been smoothed with a Gaussian function with $\sigma = 0.15 ^{\circ}$.}%
\label{tsmap}
\end{figure}

\begin{figure}
\vspace{0.4cm}
{\centering
\resizebox*{0.85\textwidth}{0.36\textheight}
{\includegraphics{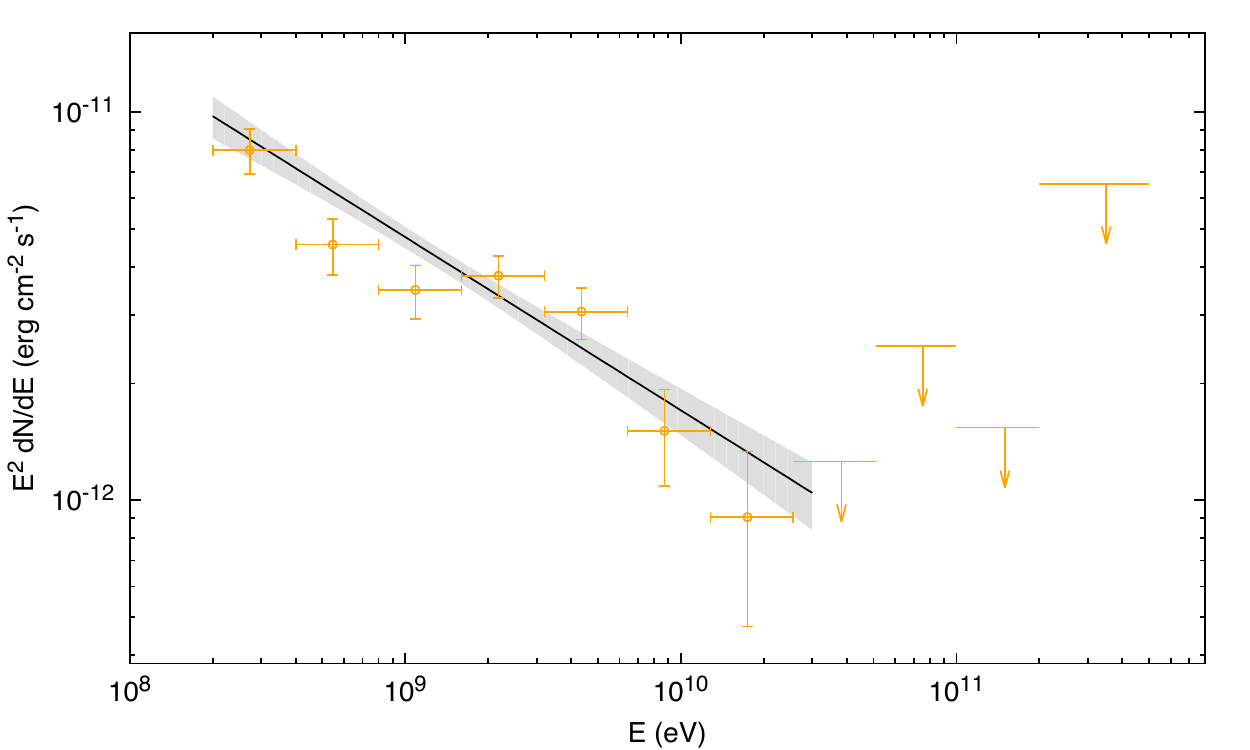}}
}
\caption{Fermi-LAT gamma-ray spectrum of the radio galaxy Cen B with  a power-law function (black line) used to fit these data.  The error bar denotes the statistical errors of 1$\sigma$.  Arrows correspond to the flux upper limits at 95\% confidence level.}%
\label{spectrum_lat}
\end{figure}

A point source located at the position of Cen B was found in the 3FGL catalog, namely 3FGL J1346.6-6027, with coordinates RA=206.652$\degree$ and Dec=-60.4537$\degree$ (J2000). This source is removed from the model in order to study the emission and obtain a new location of the source with the increased data set used here. A point source morphology was assumed for the source associated to Cen B, as searching for possible extension is outside the scope of this paper.  Events were filtered above 5 GeV to take advantage of the narrower PSF. A fit was carried out keeping the normalizations of sources located within 5$\degree$ of Cen B free. The resulting best-fit model is used to produce a background-subtracted map showing potential sources that were not accounted for by the preliminary model. Several new sources are included to improve the model of the region.  An extended region of residual emission is found at the location of the newly discovered source 3FHL J1409.1-6121e and the corresponding spatial template described in the Third Catalog of Hard Fermi-LAT Sources \citep{2017arXiv170200664T} is used to account for this emission.  An excess, $\sim$ 1.9$\degree$ from the location of Cen B, is found adjacent to the source 3FGL J1345.1-6224 (a possible supernova remnant). A new point source is added to account for this excess at the coordinates RA=205.61$\degree$ and Dec=-62.283$\degree$ (J2000), and a simple power law for the spectrum is used, in analogy to the nearby source 3FGL J1345.1-6224. The position of this source is optimized by performing a search in a grid of positions with the LAT tool gtfindsrc. The resulting TS value for this source is 32. Above 5 GeV this source and 3FGL J1345.1-6224 shows a similar spectral index, meaning that both could be part of the same extended object.    The left-hand panel in Figure \ref{tsmap} shows a TS map of the region above 5 GeV obtained by fitting a point source at each pixel location. The emission associated to Cen B is clearly seen in the map, which also shows the contours from a radio observation from the 843 MHz Sydney University Molonglo Sky Survey (SUMSS) \citep{2003MNRAS.342.1117M}.  The right-hand panel in Figure  \ref{tsmap} shows the residuals obtained after subtracting the best fit model to the data, which gives another idea of the agreement between the data and model. Emission associated to Cen B is also seen in the residuals map.
Once the model was improved with these additions, a search for the best-fit position for a source at the location of Cen B was carried out above 5 GeV with the tool gtfindsrc. The resulting coordinates were RA=206.569$\degree$ and Dec=-60.4439$\degree$ (J2000) with a 68\% confidence level error radius of 0.02$\degree$ (see Fig. \ref{tsmap}). This was basically the same location as that reported for the source associated to Cen B, 3FHL J1346.2-6026, in the Third Catalog of Hard Fermi-LAT Sources \citep{2017arXiv170200664T}, which used seven years of data above 10 GeV.\\
When the data are fitted using events above 1 GeV, the residuals obtained after subtracting the best fit model from the data are satisfactory, except for a new excess seen at the coordinates RA=207.817$\degree$ and Dec=-61.692$\degree$ (J2000).  The location of a new point source is optimized using a simple power law spectral assumption. The resulting TS value is 81 and the spectral index obtained is $2.60 \pm 0.13$. The relatively soft spectrum might explain the absence of this excess in the residuals map above 5 GeV.\\
The map of residuals shows a satisfactory modeling of all the emission in the region after adding the new point sources as well as the extended template representing 3FHL J1409.1-6121e. The final model is applied to data for the whole energy range above 200 MeV also giving satisfactory residuals.\\
In order to study the spectrum of Cen B in more detail above 200 MeV, data are divided in eleven intervals of equal size (in logarithmic scale) in energy.  A fit in each bin is done keeping the spectral parameters of all the sources fixed to the values found in a preliminary fit, except for the normalization of Cen B. It is seen that this source is not detected above $\sim 26$ GeV, which allowed us to choose the energy range from 0.2 to 26 GeV to carry out the final fit.  The spectral properties of Cen B are then explored by fitting different spectral shapes: a simple power law, a power law with an exponential cutoff and a log-parabola.   It is worth noting that at this time the normalizations of sources located within 8$\degree$ from Cen B were left free, due to the larger PSF. The resulting TS values for the different spectral shapes are, respectively, 365.5, 364.7 and 364.6, indicating that the spectrum of Cen B showed no significant spectral curvature in this range. A simple power law is adopted with a best fit spectral index of $2.45\pm 0.06$ and integrated flux of $(2.11 \pm 0.19)\times 10^{-8}$ cm$^{-2}$ s$^{-1}$. Figure \ref{spectrum_lat} shows the SED of the source with the best-fit spectral shape and its corresponding 1$\sigma$ statistical uncertainty band in the 0.2 - 26 GeV range, as well as the 95\%-confidence level upper limits obtained for the highest energy intervals where the source is not detected. The data points are obtained as indicated with a fit in each energy interval keeping the spectral index fixed to the best fit value found of 2.45. The overall source significance is 19$\sigma$.
\vspace{1.5cm}
\subsection{Time variability}
\begin{figure}
\vspace{0.4cm}
{\centering
\resizebox*{0.8\textwidth}{0.4\textheight}
{\includegraphics{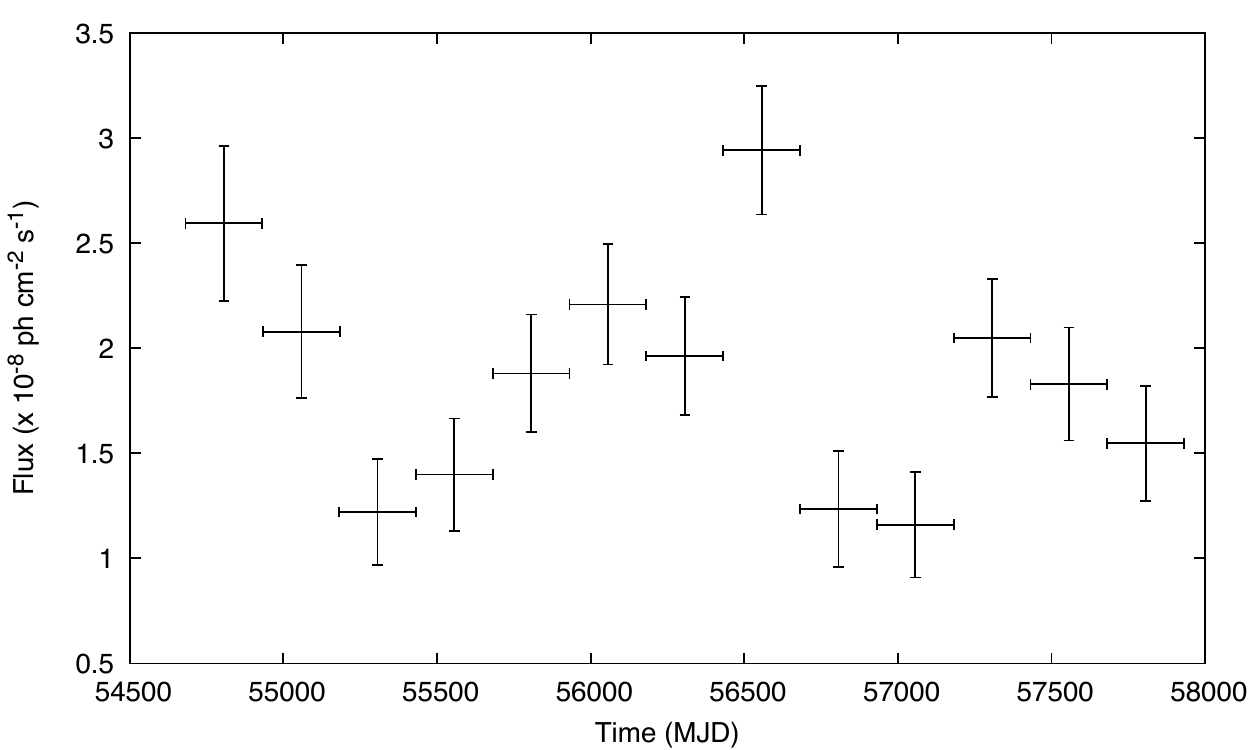}}
}
\caption{Fermi-LAT gamma-ray light curve of Cen B from 2008 August 04  (MJD 54682) to 2017 June 27 (MJD 57931).  The error bar denotes the statistical errors of 1$\sigma$.  }%
\label{lc_lat}
\end{figure}

In order to search for variability in the Fermi-LAT data, we bin the data in 13 intervals with a duration of 249 days each and calculate the source flux for each interval by fitting with the best model obtained above in the 0.2 - 26 GeV energy range. In each fit all parameters are kept fixed to the values found previously, except for the normalization of Cen B. The spectral index of a variable source could of course also change but the purpose here was to detect this variability.  This is particularly true when the statistics is relatively low in each interval and it is hard to constrain the index. It is also difficult to use shorter time intervals because of poor statistics.\\
In all the time intervals the source is detected above 4$\sigma$. The typical detection significance in the time bins lies in the range of  (6 - 7)$\sigma$. Figure \ref{lc_lat} shows the integrated flux of the source as a function of time. A variability index, $V$, is defined as a $\chi^2$ criterion as done in \cite{2010ApJS..188..405A}. The result for the variability index is $V=26.6$ when statistical errors are only used. For 12 degrees of freedom the light curve significantly deviated from that of a steady source at the 99\% confidence level for $V>26.2$. However, considering a systematic error on the flux values as low as 5\%,   $V$ reduced below this threshold, to 24.6, and we therefore do not claim that there is sufficient evidence (p-value 0.017) for variability in the $\sim 8$-month time scale studied.\\
We summarize in Table \ref{table1a} the relevant features found in the spectral analysis and the time variability after analyzing the Fermi-LAT data. In addition, the values reported in \cite{2013A&A...550A..66K} are shown. This table shows that although the uncertainties of the quantities obtained in this work are smaller due to the large data set ($\sim$ 2.5 times more cumulative Fermi-LAT data), the values reported in \cite{2013A&A...550A..66K} are compatible with our results.

\begin{table}[ht!]\renewcommand{\arraystretch}{1.45}\addtolength{\tabcolsep}{0.6pt}
\centering
\caption{Summary of the relevant features in the Fermi-LAT data}\label{table1a}
\begin{tabular}{ l c c }
 \hline
                                                                      &        This work                                            & Other work$^a$      \\
 \hline 
 \hline
\scriptsize{Analyzed interval}              & \scriptsize{August 2008 - June 2017}  & \scriptsize{August 2008 - February 2012}     \\
\scriptsize{Spectral index }                              & \scriptsize{2.45 $\pm$ 0.06}  & \scriptsize{2.6 $\pm$ 0.2}     \\
\scriptsize{Flux\, (cm$^{-2}$ s$^{-1}$) }          & \scriptsize{$(2.11 \pm 0.19)\times 10^{-8}$}  &  \scriptsize{$(1.9 \pm 0.5)\times 10^{-8}$}     \\
\scriptsize{Variability }                                     & \scriptsize{No found}  &  \scriptsize{No found}     \\

\hline
\end{tabular}

\begin{flushleft}
\scriptsize{ $^a$ Analysis shown in \cite{2013A&A...550A..66K}}
\end{flushleft}

\end{table}
\section{Modeling the Spectral Energy Distribution}

\subsection{The extended  lobes}
\begin{figure}
{\centering
\resizebox*{1\textwidth}{0.65\textheight}
{\includegraphics{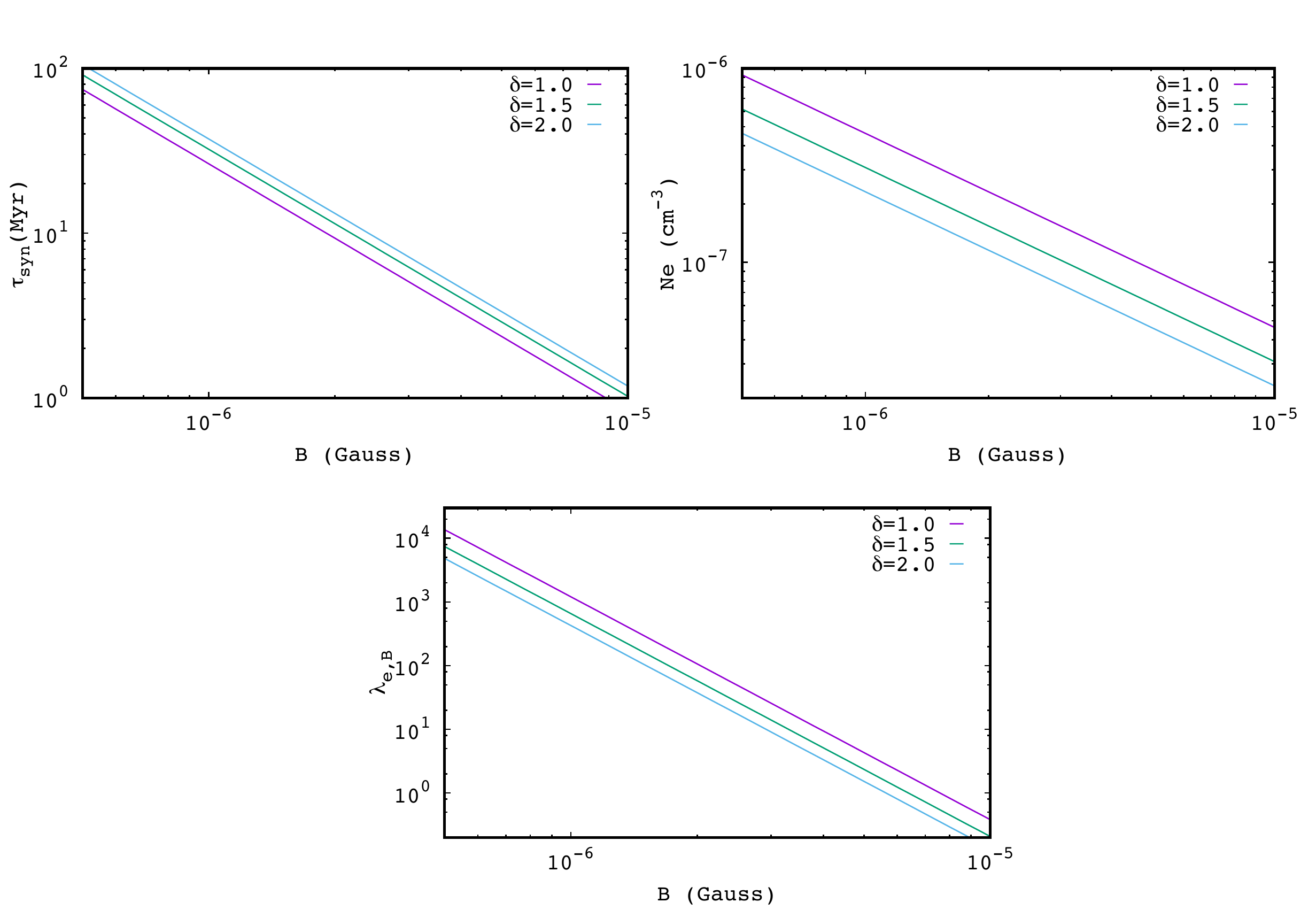}}
}
\caption{The best-fit values of parameters $\tau_{\rm syn}$, $N_e$ and $\lambda_{e,B}$ as a function of $B$ and $\delta_D$ obtained after describing the radio wavelength data with synchrotron radiation.} 
\label{parameters}
\end{figure}

\begin{figure}
{\centering
\resizebox*{0.7\textwidth}{0.32\textheight}
{\includegraphics{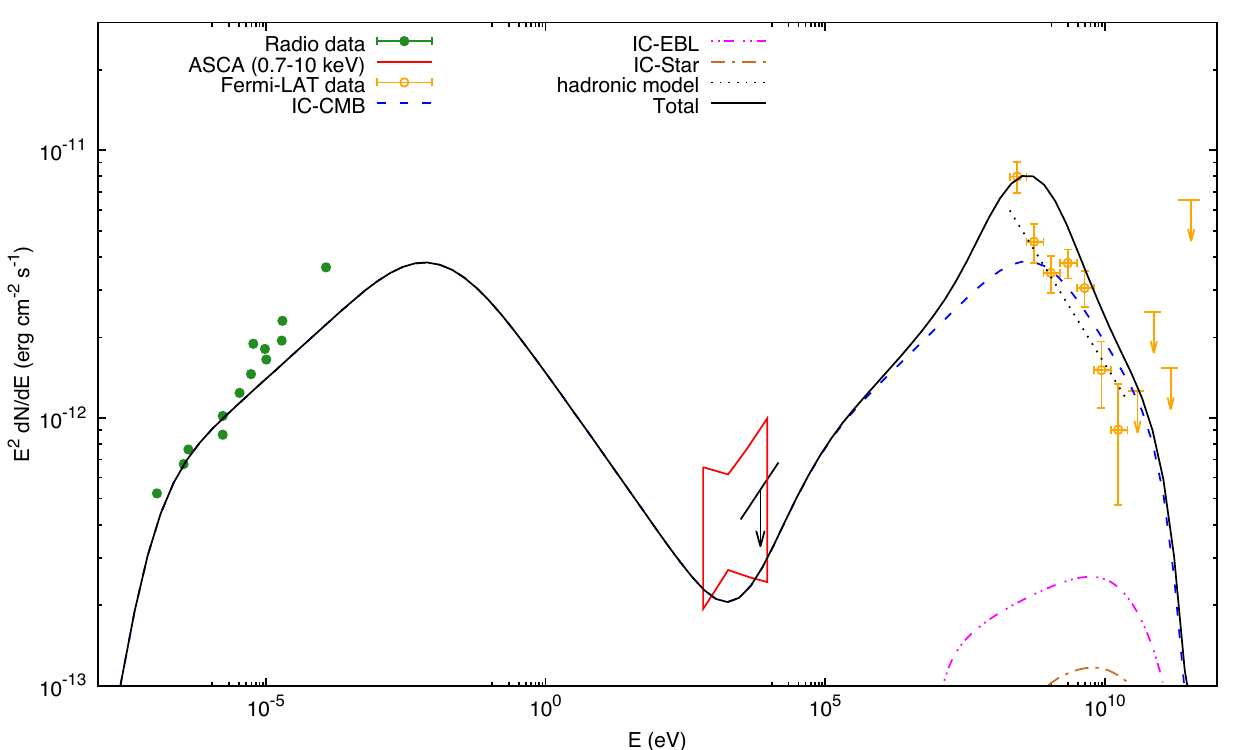}}
}
\caption{The broadband SED of the extended  lobes of Cen B with the model curves for the synchrotron, external inverse Compton with CMB, EBL and the starlight of the host galaxy and hadronic emissions.  Synchrotron radiation model is used to  describe the radio wavelength data and a superposition of external inverse Compton with CMB (dashed blue line), EBL (double dotted-dashed magenta line)  and the starlight (dashed-dotted brown line) and  hadronic models (dotted black line)  are used to fit the gamma-ray spectrum.   The thick black curve denote the total emission.   The  best-fit values of our lepto-hadronic  model are reported in Table \ref{values_lobe}. The radio fluxes (green full circles) are taken from  \cite{2001MNRAS.325..817J},  \cite{2004AJ....128.2593F}  and \cite{2009MNRAS.395..504B}. The red bow tie and the thick black line correspond to the ASCA measurements  \citep{1998ApJ...499..713T}  and the Suzaku upper limit \citep{2013A&A...550A..66K} in the energy ranges of 0.7 - 10 keV and 2 - 10 keV, respectively. The Fermi-LAT data shown by gold open circles correspond to the analysis displayed in Section \ref{sec:2}}
\label{sed_lobe}
\end{figure}

\begin{figure}
{\centering
\resizebox*{0.8\textwidth}{0.35\textheight}
{\includegraphics[angle=90]{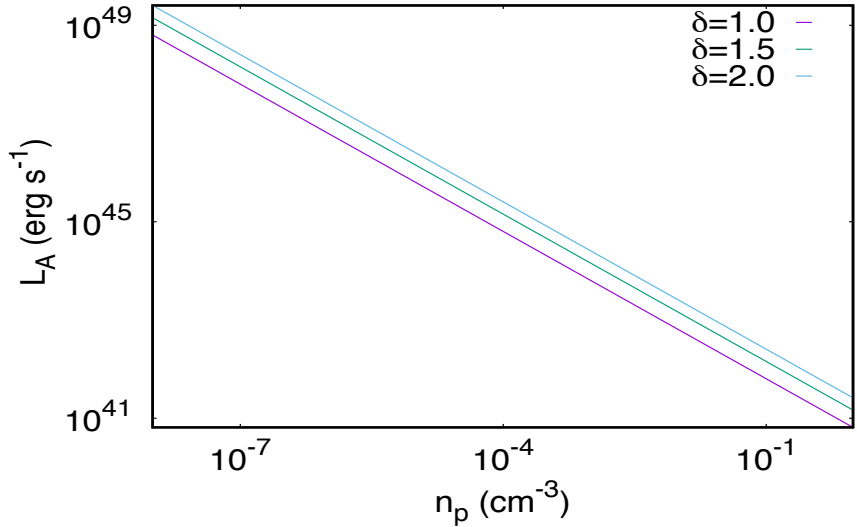}}
}
\caption{CR luminosity as a function of the target proton density within the extended lobes.  The CR luminosity is normalized using the gamma-ray spectrum and the $\pi^0$ decay products generated in the hadronic interactions.}
\label{proton_luminosity}
\end{figure}

The radiative processes used to interpret the broadband SED of the extended  lobes are shown in  appendix  A.  The synchrotron model is used to describe the radio wavelength data.  The best-fit values of synchrotron spectral breaks (the characteristic $\epsilon^{\rm syn}_{\gamma,m}$ and cooling $\epsilon^{\rm syn}_{\gamma,c}$  break energies) and the proportionality constant of the synchrotron spectrum  ($A^{\rm syn}_{\gamma}$) were obtained using the Chi-square ($\chi^2$) minimization method implemented in the ROOT software \citep{1997NIMPA.389...81B,2017ApJS..232....7F, 2016ApJ...826...31F}.  Using the eq. (\ref{synrad}), we plot in Figure \ref{parameters} the equipartition parameter, the electron number density and the synchrotron timescale as a function of the magnetic field for three different Doppler factors $\delta_D$=1.0, 1.5 and 2.0. Considering the magnetic field in the range of  ($5\times 10^{-7}$ - $10^{-5}\,$) G for $\delta_D$=1.5,  the cooling timescale, the electron number density and the equipartition parameter lie in the range of (1 - 100) Myr, ($4\times10^{-8}$ - $9\times10^{-7}$)\,${\rm cm^{-3}}$ and  ($5\times10^{-1}$ - $5\times10^4$), respectively.    For instance, considering the values of the magnetic field $B\simeq 3.8\,{\rm \mu G}$ and the Doppler factor  $\delta$=1.5, the cooling timescale, electron number density and the equipartition parameter are   $\tau_{\rm syn}\simeq$ 4.1 Myr, $N_e\simeq1.6\times 10^{-7}\,{\rm cm^{-3}}$ and $\lambda_{e,B}\simeq$ 4.7, respectively.   This analysis leads to an estimate of the minimum electron Lorentz factor of $\gamma_{\rm e,m}\simeq$ 3.   The age of the  lobes can be estimated through the cooling timescale of the radiating electrons $ \tau_{\rm lobe}\approx \tau_{\rm syn}$= 4.1 Myr. Using the best-fit values for $\delta_D=1.5$, we calculate the magnetic and electron luminosities which are  given by {\small $L_B\simeq 9.0\times 10^{44}\,\,{\rm erg\,\, s^{-1}\, \left(\frac{R}{50\, {\rm kpc}}\right)^2\,  \left(\frac{B}{\rm 5\mu G}\right)^2\, \left(\frac{\Gamma}{1}\right)^2}$} and {\rm $L_e\simeq 4.5\times 10^{45}\,\,{\rm erg\,\, s^{-1}\left(\frac{\lambda_{e,B}}{5}\right)\left(\frac{R}{50\, {\rm kpc}}\right)^2\,  \left(\frac{B}{\rm 5\mu G}\right)^2\, \left(\frac{\Gamma}{1}\right)^2}$} , respectively.  The quantities $R$ and $\Gamma$ are the size of the extended  lobes and the bulk Lorentz factor, respectively.\\
\\
The leptonic and hadronic models shown in appendix  A are used to describe the Fermi-LAT data. We consider the effect of EBL absorption modelled in  \cite{2008A&A...487..837F}.  For the leptonic model,  the contribution of synchrotron and external inverse Compton (IC) emissions from photon fields of  cosmic microwave background  (CMB), extragalactic background light (EBL)  and the starlight of the host galaxy are considered.  The values of the frequencies and the energy densities of the photon fields associated to CMB, EBL and the starlight of the host galaxy are taken from \cite{2013SAAS...40..225D, 1977A&A....59L...3L,2007A&A...466..481S}.   For the hadronic model, we use the neutral pion ($\pi^0$) decay products from the hadronic interactions between ultra-relativistic CRs and the target proton density within the extended  lobes.  Figure \ref{sed_lobe} shows the broadband SED  of the extended  lobes of Cen B with the best-fit curve smoothed with the sbezier function\footnote{gnuplot.sourceforge.net.}  and obtained from the leptonic and hadronic contributions.  In this figure is shown that the emission generated by the IC-CMB and the hadronic models can account for the gamma-ray flux, and also that the contribution of the  IC-EBL and the IC-star models to the gamma-ray flux is too small.   Using the best-fit values of the hadronic model, we normalize the CR luminosity as a function of the target proton density, as shown Figure \ref{proton_luminosity} . This figure displays that the CR luminosity lies in the range of $10^{41}$ - $10^{49}$ erg ${\rm s^{-1}}$ for a proton density in the range of ($10^{-8}$ -  $10^{-1}$)    $\,\,{\rm cm^{-3}}$. The  derived parameters from the CR interactions such as non-thermal pressure, proton density energy and total energy  are reported in Table \ref{values_lobe}. 
%
%
\subsection{The region close to the core}
\begin{figure}
{\centering
\resizebox*{0.8\textwidth}{0.32\textheight}
{\includegraphics{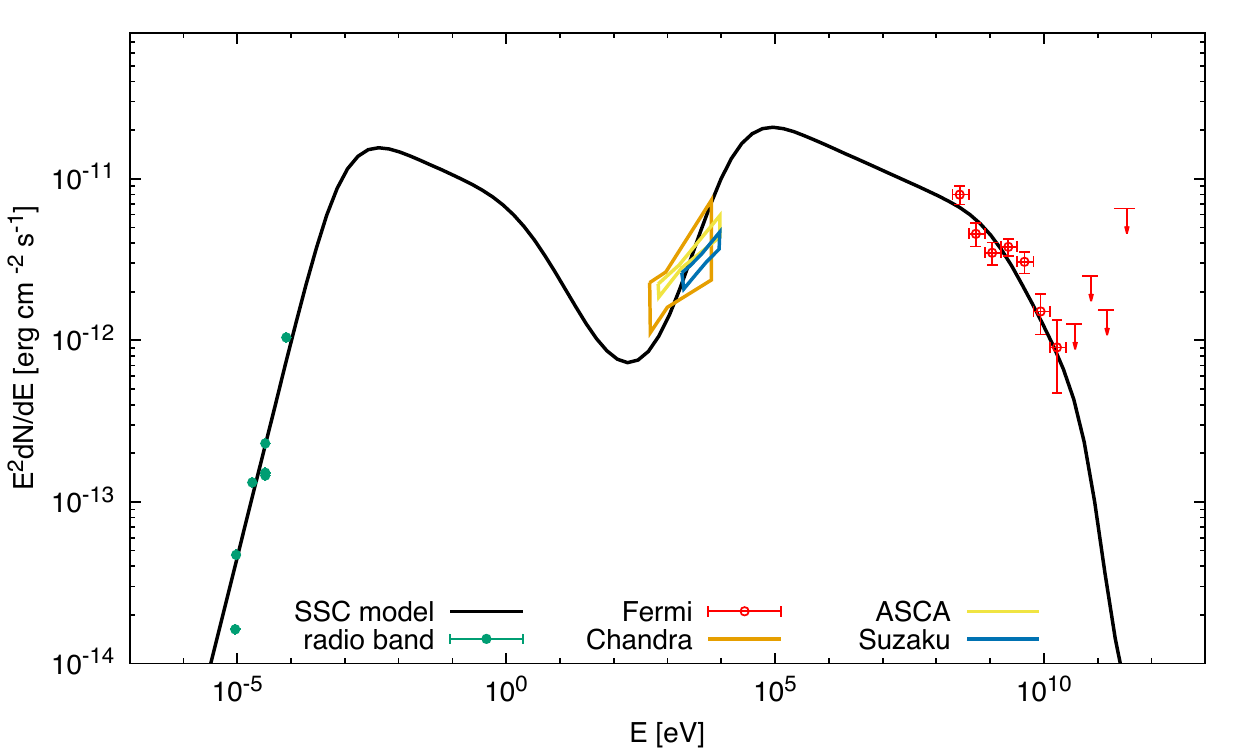}}
}
\caption{The broadband SED of the core of  Cen B with the one-zone SSC model fit (thick black line) obtained with the model presented in \cite{2016ApJ...830...81F}.  The best-fit parameters are reported in Table \ref{values_core}.  The radio fluxes (green full circles) are taken from  \citep{2001MNRAS.325..817J}.  The brown, gold and blue bow ties correspond to Chandra 0.5 - 7 keV \citep{2005ApJS..156...13M},  ASCA  0.7 - 10 keV \citep{1998ApJ...499..713T}  and  Suzaku 2 - 10 keV \citep{2013A&A...550A..66K}  measurements.  The Fermi-LAT data shown by red open circles correspond to the analysis displayed in Section \ref{sec:2}. }
\label{sed_core}
\end{figure}

The most accepted model to describe the broadband SED from relativistic jets in radio galaxies is the one-zone SSC model \citep{2010ApJ...719.1433A, 2014MNRAS.441.1209F}.  In this scenario, accelerated electrons are confined by magnetic fields in the emitting region.  This region is moving at the relativistic velocities along the jet.  Low-energy photons are emitted by synchrotron radiation and scattered up to higher energies by inverse Compton scattering.  Photons detected from radio to optical bands are described with synchrotron emission and from X-ray to gamma-ray bands are explained via inverse Compton emission.  In order to interpret the broadband SED detected close to the core of Cen B, we use the one-zone SSC model showed in \cite{2016ApJ...830...81F}.\\
The Chi-square $\chi^2$ minimization routine in the ROOT software  \citep{1997NIMPA.389...81B} is used to obtain the best-fit curve with the values of  the Doppler factor, the size of emitting region, the electron number density and the magnetic field.  In addition, the sbezier function implemented in the gnuplot software is used to smooth the best-fit curve.\\
Figure \ref{sed_core} shows the broadband SED emitted from the region close to the core with the best-fit model curve.   This figure shows that the one-zone SSC model generated by the electron population is successful in describing the SED. The best-fit parameters obtained with our model are reported in Table \ref{values_core}.  Using the best-fit values,  several physical properties such as the total jet power and the energy densities carried by electrons and magnetic field are derived.  This table displays all the parameter values obtained and derived in and from the fit for the values of the distance $d_z=56$ Mpc,  the viewing angle $\theta \approx$ 20$^\circ$ and the minimum electron Lorentz factor $\gamma_{\rm e,min}$ = 100.    The values shown in Table \ref{values_core} are in the range of those reported by Katsuta et al. \cite{2013A&A...550A..66K}.  The ratio between the magnetic field and electron density of $U_e/U_B\simeq$ 200 indicates that these quantities cannot be associated to a mechanism of energy equipartition.  Considering the values of the Doppler factor and the size of emitting region, we found that the variability time scale  is $\sim$ one month.  The CR luminosity was estimated through the charge neutrality condition in the jet  (N$_e$=N$_A$) \citep{2013ApJ...768...54B, 2009ApJ...704...38S, 2011ApJ...736..131A, 2014A&A...562A..12P,2017APh....89...14F}. In this case, the CR luminosity becomes $\simeq1.3\times 10^{46}$ erg/s.  Considering a typical value of a supermassive black hole associated to the radio galaxies ($\sim 10^8\,M_{\odot}$), then  the value of the CR luminosity estimated corresponds only to a fraction of the Eddington luminosity. The best-fit value found of the size of the emitting region  $1.7 \times 10^{17}$ cm indicates that it is far away from  the typical gravitational radius which usually lies in the range of ($10^{13}$ - $10^{14}$) cm.
\section{UHE cosmic rays}
Radio galaxies have been amply proposed as potential candidates to accelerate CRs up to UHEs  via  Fermi mechanisms \citep{2009NJPh...11f5016D,2018arXiv181010294K},  magnetic reconnection \citep{ 2016MNRAS.455..838K, 1987ApJ...315..504L, 2013MNRAS.430.2828L},  high temperature gradients  \citep{2013A&A...558A..19W} and magnetized accretion disks working as electric dynamos \citep{2005Ap&SS.298..115L, 1976Natur.262..649L}.\\
Data collected by PAO between  2004 January 1$^{\rm th}$ and 2007 August 31  exhibited the arrival directions of 27 UHECRs  with energies larger than $5.7 \times 10^19$ eV \citep{2007Sci...318..938P, 2008APh....29..188P}.   A statistically significant correlation was claimed  between these UHECR events and the distributed galaxies reported in the $12^{\rm th}$ Veron-Cetty-Veron (VCV) catalog of quasars and active nuclei \citep{2006A&A...455..773V}.   The correlation was highly significant within $3.1^\circ$ of nearby AGN located inside 75 Mpc  (z $<$ 0.018 ).\\
The PAO dataset exhibited  a clustering of at least one and three events in a region of 5$^\circ$ around  the radio galaxies Cen B and Cen A, respectively  \citep{2007Sci...318..938P, 2008APh....29..188P,2009ApJ...693.1261M}.  
This result has triggered an extensive of interest of explaining mechanisms of acceleration and places  where UHECRs can be accelerated  inside  these radio galaxies. It has also carried to forecast studies of high-energy neutrinos and gamma-rays.\\
Recently,  Aab et al. \cite{2015ApJ...804...15A} reported a dataset of 120 events collected during 10 years of operations (from 2004 January 1$^{th}$ to 2014 March 31).  Aab et al. \cite{2014PhRvD..90l2006A}  studied the composition of the UHECR events in the dataset collected between December 2004 and December 2012.  This collaboration used  three hadronic interaction models in order to fit the chemical  composition of UHECRs in the entire energy range.   They reported that the dataset was well-fitted with a mix of protons or iron nuclei only at energies less than $\lesssim10^{19}$ eV while light nuclei such as  Helium and Nitrogen described successfully dataset at the highest energies.  Fraija et al. \cite{2018MNRAS.481.4461F} associated 14 UHECRs in a region of 15$^\circ$ around Cen A and showed that these events could be accelerated inside the giant  lobes of Cen A. In addition, authors studied the composition of these events and found that  the most promise candidates of UHECR composition are light nuclei such as Carbon/Nitrogen nuclei.  In this section, a similar analysis will be done for Cen B.\\
\subsection{Chemical composition} 
\begin{figure}
\vspace{0.4cm}
{\centering
\resizebox*{0.9\textwidth}{0.47\textheight}
{\includegraphics{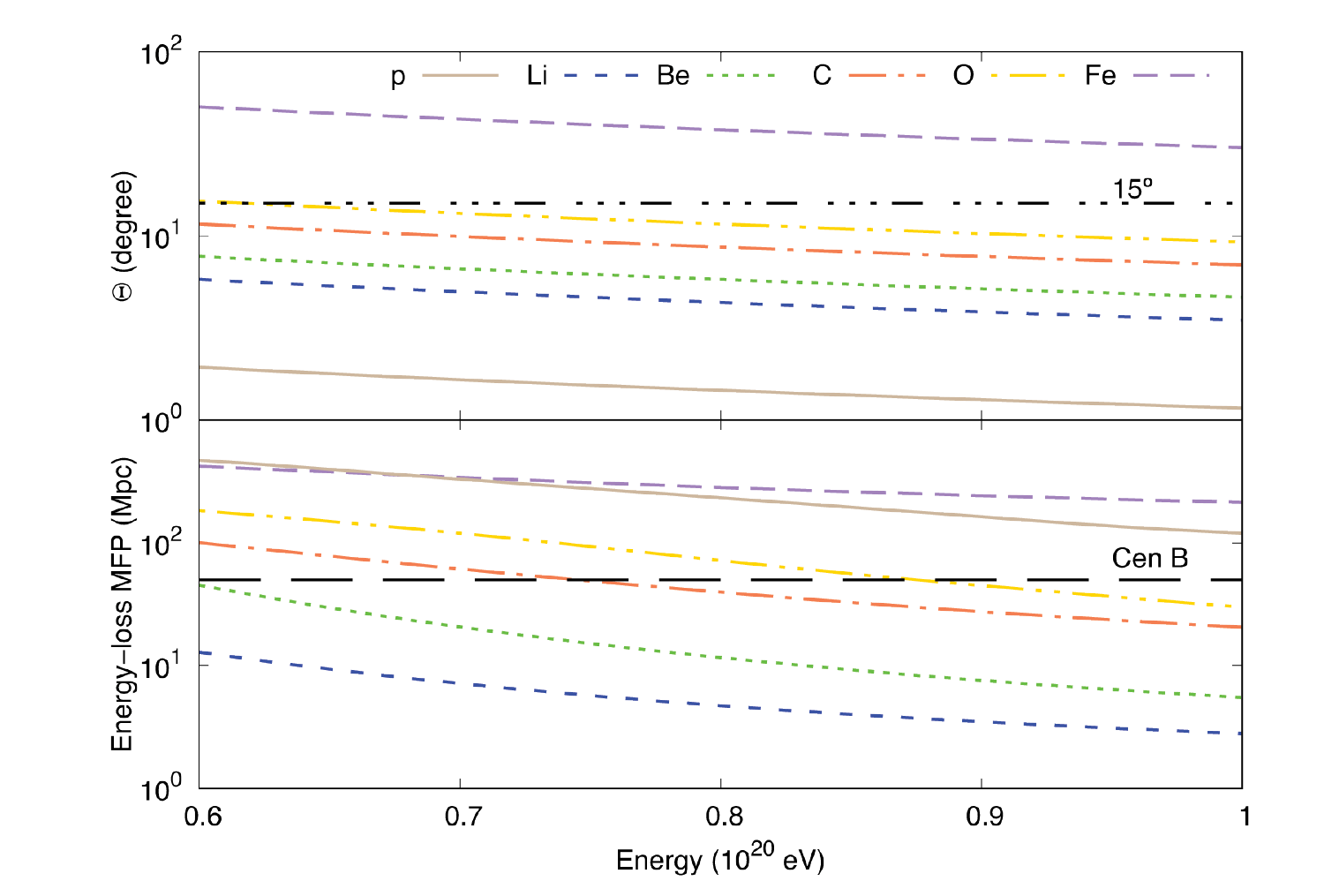}}
}
\caption{The average deflecting angle (upper)  and  the mean free path of photo-disintegration  processes (lower) as a function of energy for  UHE protons (p) and nuclei (Li, Be, C, O and Fe).}%
\label{mfp}
\end{figure}

The radiation and the magnetic fields play an important role in the chemical composition of UHECRs.  The large structure of the Universe and distinct extragalactic and galactic configurations of magnetic fields (i.e. galactic winds, magnetic winds, magnetic turbulences) make it possible for UHECRs to interact  along their paths with radiation fields and also to be deviated from their birthplaces.  For heavy nuclei,  larger deflections are expected, both from the extragalactic and Galactic magnetic fields.  UHECRs interact with  CMB and EBL (infrared to ultraviolet background) photon fields. Because of the efficiency of the hadronic interactions depends on the amount of target photons,  UHECRs primarily interact with CMB than EBL photons\footnote{The photon density of CMB is  2 to 4 orders of magnitude larger than the photon density of EBL.}.  If UHECRs are accelerated within Cen B,  their chemical composition in a good approximation could be estimated through the photo-disintegration processes and the deviations due to extragalactic and Galactic magnetic fields.\\
The mean free path due to photo-disintegration processes in the $\delta$-function is  \citep{2008JPhCS.120f2006D}
{\small
\be\label{lambdaE}
\lambda(E,A)\simeq 2.5\,\,{\rm Mpc}\, \left(\frac{E_A}{5.8\times10^{19}\,{\rm eV}}\right)^2\, \frac{\exp\left\{{\frac{0.83A^{0.79}\,}{(1+z)}\left(\frac{5.8\times 10^{19}\,{\rm eV}}{E_A}\right) } \right\}} {k\, A^{1.79}(1+z)}\,,
\ee
}
where  $E_A$ is the threshold energy of the nuclei ``${\rm A}$" and ``${\rm k}$"  is 1.2, 3.6 and 4.349 for A=4, 10 $\leq A \leq $ 22 and 23 $\leq A \leq $ 56, respectively. Similarly, UHE protons in their paths interact primarily with CMB photons.   The mean free path of UHE protons due to photopion energy losses is \citep{2007arXiv0711.2804D, 2009herb.book.....D}
{\small
\be\label{lambdap}
\lambda_{p}(E_p)\simeq 13.6\,\,{\rm Mpc}\,\,\frac{ \exp\left\{13\left(\frac{5.8\times10^{19}\,{\rm eV}}{E_p}\right)\right\}}{1+  13\left(\frac{5.8\times10^{19}\,{\rm eV}}{E_p}\right)}\,.
\ee
}
The lower panel in Figure \ref{mfp} shows the mean free path of UHE nuclei and protons for photo-disintegration processes  and photopion energy losses due to CMB photons, respectively.  The dashed line represents the distance between Earth and Cen B, the solid lines in blue, green, orange, yellow and purple colors are Lithium (Li), Beryllium (Be),  Carbon (C), Oxygen (O) and Iron (Fe) nuclei, respectively, and the solid line in brown color corresponds to protons (p).   In this figure can be inferred that a degree of anisotropy in the chemical composition of UHECRs could be expected from a source  located at 56 Mpc, i.e.  Cen B.  In this case,  nuclei heavier than Carbon ($Z \geq 6$) have mean free paths substantially larger than the distance of Cen B and therefore, these nuclei can be expected. Otherwise,  the lightest nuclei with $Z \leq 5$ can hardly  reach Earth. \\
\\
UHECRs are deflected by magnetic fields along their trajectories in the extragalactic and Galactic medium. Although significant progresses have been made to describe the Galactic magnetic field  in the last two decades \citep[see e.g.,][]{1999MNRAS.306..371H,  1999JHEP...08..022H, 2002APh....18..165T,  2008NuPhS.175...62H},  only minor advances have been made to understand the extragalactic magnetic field.\\ 
Whereas the strength of Galactic magnetic field is estimated through the pulsar rotation and the distribution of free electrons \citep{2003astro.ph..1598C, 2002astro.ph..7156C},  its orientation is inferred mainly through the description of radio and gamma-ray emission.   It is known that the inhomogeneities of the Galactic magnetic field cause distinct angular deflections depending on the direction of UHECR propagations \citep{1999JHEP...08..022H, 2007APh....26..378K, 2008ApJ...681.1279T}.  On the plane, the magnetic field is twice to three times stronger  than in the halo, and also UHECRs in their trajectories towards Earth would  have to cross twice the size of our galaxy ($L_G$) \citep{2008PhyS...78d5901F}. The average deflecting angles for nuclei with charge Z and energy $E_{\rm A}$ are \citep{2008PhyS...78d5901F}
{\small
\be
\theta_{\rm A,G}\sim  1.3^\circ\, Z \, \left(\frac{E_{\rm A}}{5.8\times10^{19}\, {\rm eV}}\right)^{-1}\,  \left(\frac{B_G}{1\,\mu G}\right) \sqrt{\frac{L_G}{10\,{\rm kpc}}}\,\sqrt{\frac{l_{\rm c,G}}{{\rm kpc}}}\,, 
\ee
}
in the halo, and 
{\small
\be
\theta_{\rm A,G}\sim  1.9^\circ\, Z \, \left(\frac{E_{\rm A}}{5.8\times10^{19}\, {\rm eV}}\right)^{-1}\,  \left(\frac{B_G}{4\, \mu G}\right) \sqrt{\frac{L_G}{20\,{\rm kpc}}}\,\sqrt{\frac{l_{\rm c,G}}{{\rm kpc}}}\,, 
\ee
}
on the plane. Here,  $l_{\rm c,G}$ is the coherence length of the Galactic magnetic field.\\ 
The orientation and strength of the extragalactic magnetic field definitely alter the propagation of UHECRs.  Using Faraday rotation measurements, magnetic fields in the core and outside of clusters of galaxies  have been estimated \citep{1998A&A...335...19R, 1999ApJ...514L..79B}.  Additionally, Neronov and Semikoz \cite{2009PhRvD..80l3012N} proposed that the analysis of the extended gamma-ray emission around point sources together with time delays during flaring gamma-ray activities  could provide strong constraints of the extragalactic magnetic field. These studies led to an estimate of the extragalactic magnetic field in the range of ($10^{-16}\, -\, 10^{-9}$)  ${\rm G}$. Therefore, the maximum 
 mean-square deviation can be estimated as
{\small
\be
\theta_{\rm A, EG}\sim  0.6^{\circ}Z\left(\frac{5.8\times10^{19}\, {\rm eV}}{E_{\rm A}}\right)^{-1} \left( \frac{B_{\rm EG}}{{\rm 1\,nG}} \right)   \sqrt{\frac{L_{\rm EG}}{56\, {\rm Mpc}}}  \sqrt{\frac{l_{\rm c,EG}}{1\, {\rm Mpc}}}\,,
\ee
}
where  $l_{\rm c, EG}$  is the coherence length of the extragalactic magnetic field.     Considering the location of Cen B ($\sim$ 2$^\circ$ from the galactic plane and  $\sim$ 42$^\circ$ from the galactic center),  UHECRs would be deflected with larger angles due to the Galactic magnetic field than the extragalactic one. \\
The upper panel in Figure \ref{mfp} shows the largest deflecting angle of UHE nuclei and protons as a function of energy.  The largest deflecting angle is computed considering the Galactic magnetic field  in the halo and the extragalactic magnetic field.  In this figure can be seen that a degree of anisotropy in the chemical composition of UHECRs is expected if we consider a radius of $\sim 15^\circ$ centered in Cen B.   In this case, the UHE nuclei for $Z > 8$ are deflected during their trajectories  by more than $15^\circ$ in the Galactic and extragalactic magnetic fields, and therefore they  could not be expected from Cen B.\\ 
Taking into the consideration both the photo-disintegration processes due to CMB photons and the largest deflecting angles induced by the Galactic and extragalactic magnetic fields,  a large fraction of protons and light nuclei such as Carbon, Nitrogen and Oxygen are expected from this radio galaxy.    
\subsection{Condition and Timescales}
In order to discuss if Cen B can accelerate CRs up to energies as high as $10^{20}\,{\rm eV}$, we use the Hillas criterion with the best-fit values obtained from the description of the broadband SED.  The Hillas criterion provides a bound to the maximal energy that CRs can reach inside a confined region of size $R$ and magnetic field $B$ \citep{1984ARA&A..22..425H}. This criterion, although constitutes a necessary condition for accelerating CRs up UHEs,  under no circumstances can be regarded as a sufficient condition. This criterion must be analyzed together with additional limitations due to the radiative losses and the timescales when CRs diffuse in the entire magnetized region. In fact, the acceleration timescale must be less than the escape timescale and the age of the source. The maximal energy that CRs can reach depends on the orientation and strength of the magnetic fields in the relativistic shocks.  Reville and Bell \cite{2014MNRAS.439.2050R} suggested that to elude this limitation, a highly disorderly field is needed on larger scales.   Although, magnetic  fields immersed on short scales play an important role  in Fermi processes at relativistic shock waves, it hinders the acceleration efficient of CRs \citep{2014MNRAS.439.2050R, 2010MNRAS.402..321L}.   Another ingredient that makes this criterion more rigorous is associated with the acceleration processes in a non-relativistic regime. In this case, the velocity $\beta$ has to be considered, thus limiting the maximal energy. \\
\\
Considering the Hillas criterion  $R>2 R_g/\beta$ with $R_g=\frac{E_{\rm A}}{eB}$ the Larmor radius \citep{1984ARA&A..22..425H},   the maximal energy becomes 
{\small
\be\label{Emax}
E_{\rm A,max}\simeq  2.6\times 10^{19}\,{\rm eV}  Z \,\left(\frac{\beta}{0.1} \right)\,\left(\frac{B}{3\times 10^{-6}\, {\rm G}} \right)\, \left(\frac{R}{100\, \rm kpc} \right)\,.  
\ee
}
Considering the energy range observed by PAO $\sim 5.8 - 10 \times 10^{19}$ eV and the Hillas criterion (eq. \ref{Emax}),  we conclude that protons are excluded as possible CRs detected by  PAO from the radio galaxy Cen B.   The values of $\beta =0.1- 0.2$ \citep{2009NJPh...11f5016D} and the magnetic field of $B=3\times 10^{-6}$  G \citep{1998ApJ...499..713T} in a scale of $R\simeq 100$ kpc (5 arcmin) \citep{2001MNRAS.325..817J} are consistent with the condition necessary so that CRs can be accelerated up to energies larger than $10^{19}$ eV  \citep{2014MNRAS.439.2050R}.  \\ 
The acceleration  timescale  $t_{\rm A,acc}\simeq \eta R_g$ in the  lobes can be obtained by
{\small
\be\label{tac}
t_{\rm A, acc}\simeq 3.5\, {\rm Myr}\,\,Z^{-1}\,\eta \left(\frac{E_{\rm A,max}}{10^{19}\,{\rm eV}}\right)\left(\frac{B}{3\times 10^{-6}\,{\rm G}} \right)^{-1}\, \left(\frac{\beta}{0.1} \right)^{-2}     ,    
\ee
}
where $\eta$ is the gyromagnetic factor.   CRs within  lobes can diffuse by  $t_{\rm A, esc}\simeq \frac{R^2}{2\,D(E)}$, with $D(E)=\frac13\eta R_g$ the diffusion coefficient. Hence, the  escape timescale is given by 
{\small
\be\label{tdif}
t_{\rm A, esc}\simeq  4.6\, {\rm Myr}\,Z\,\eta^{-1}   \left(\frac{B}{3\times 10^{-6}\,{\rm G}} \right) \left(\frac{R}{100\, \rm kpc} \right)^2 \left(\frac{E_{\rm Z,max}}{10^{19}\,{\rm eV}}\,\right)^{-1}\,.
\ee
}
By comparing the acceleration and escape timescales can be seen that $t_{\rm A, acc}\simeq t_{\rm A, esc}\ll t_{lobe}$.\\
\\
On the other hand,  a similar analysis can be done close to the core using  the best-fit values of the parameters obtained and derived from the fit of the broadband SED  (see  Table \ref{values_core}). Taking into consideration the values of the bulk Lorentz factor, the magnetic field and the size of emitting region, the maximum energy that CRs can reach is $3.59\times 10^{19}$ eV. Therefore,  ultra-relativistic protons cannot be accelerated to energies much higher than few tents of EeV. 
\vspace{0.5cm}
\subsection{Statistical Analysis}\label{statistical}
\begin{figure}
\vspace{0.4cm}
{\centering
\resizebox*{0.9\textwidth}{0.4\textheight}
{\includegraphics{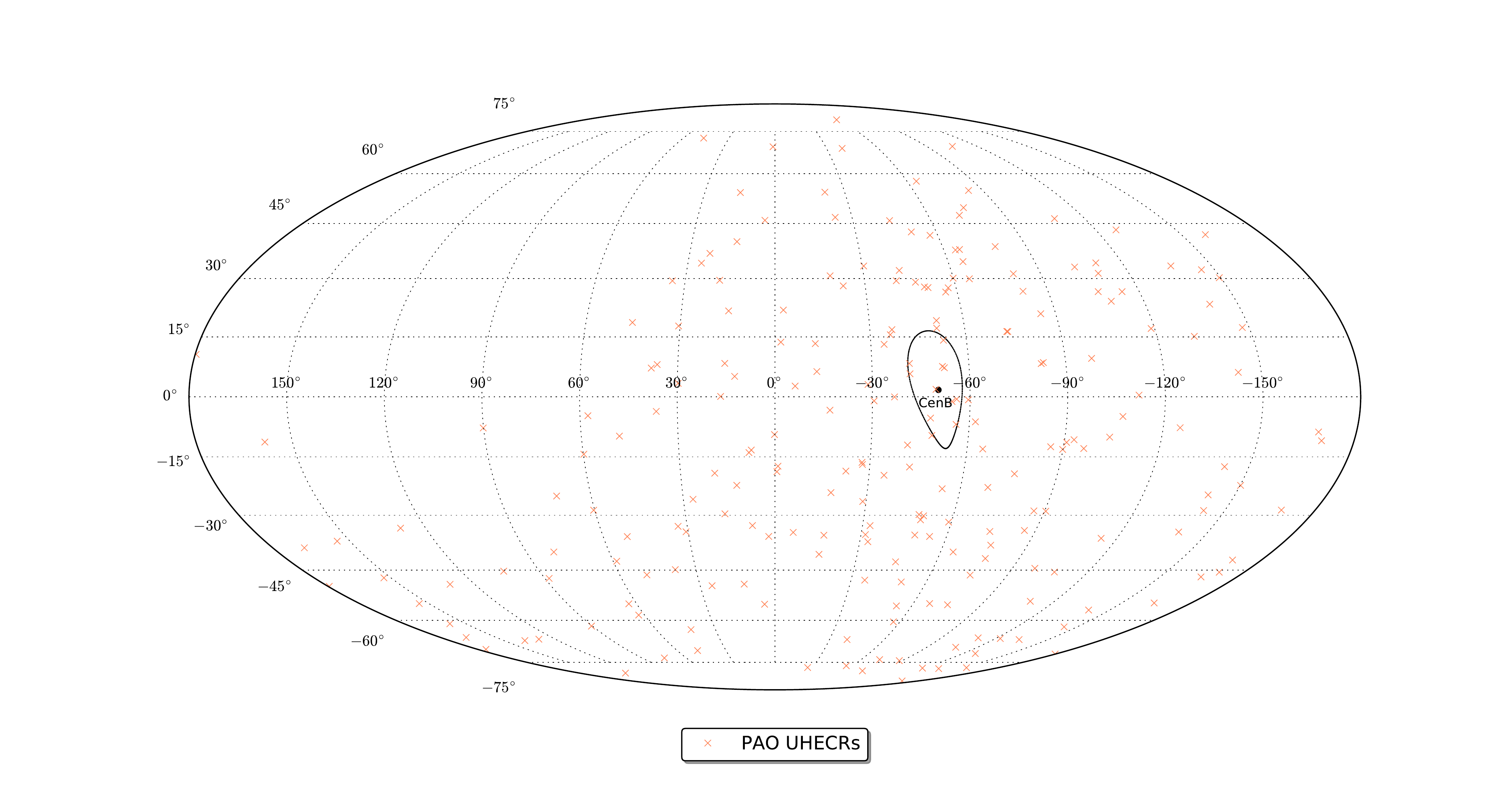}}
}
\caption{Sky-map (in Galactic coordinates) of UHECRs with E$\geq 5.8\times10^{19}\, {\rm eV}$ collected by PAO (red crosses) over 10 years of operations, from 2004 January 1$^{\rm th}$ up to 2014 March 31 \citep{2015ApJ...804...15A} and a black circle of 15$^\circ$ radius around the direction of Cen B, indicated as black point.}
\label{Skymap_cenB}
\end{figure}

\begin{figure}
\vspace{0.4cm}
{\centering
\resizebox*{0.6\textwidth}{0.4\textheight}
{\includegraphics{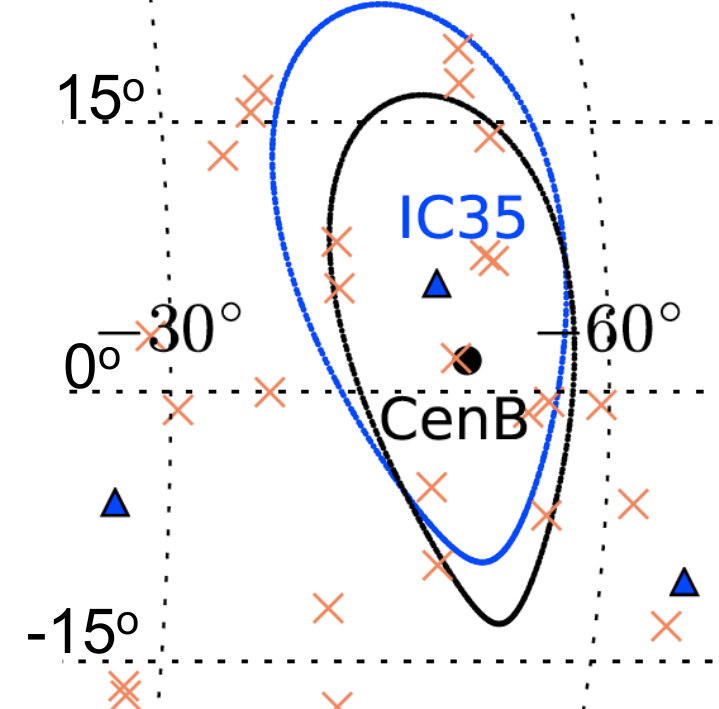}}
}
\caption{A blow-up of the sky-map (in Galactic coordinates) around Cen B (black point) with a black circle of 15$^\circ$ around this radio galaxy  is shown.  UHECRs and HE neutrinos are displayed in red crosses and blue triangles, respectively. The blue contour represents the median angular error of IC35 neutrino event.}
\label{SkymapZoom}
\end{figure}

\begin{figure}[hbt]
  \includegraphics[height=0.8\linewidth]{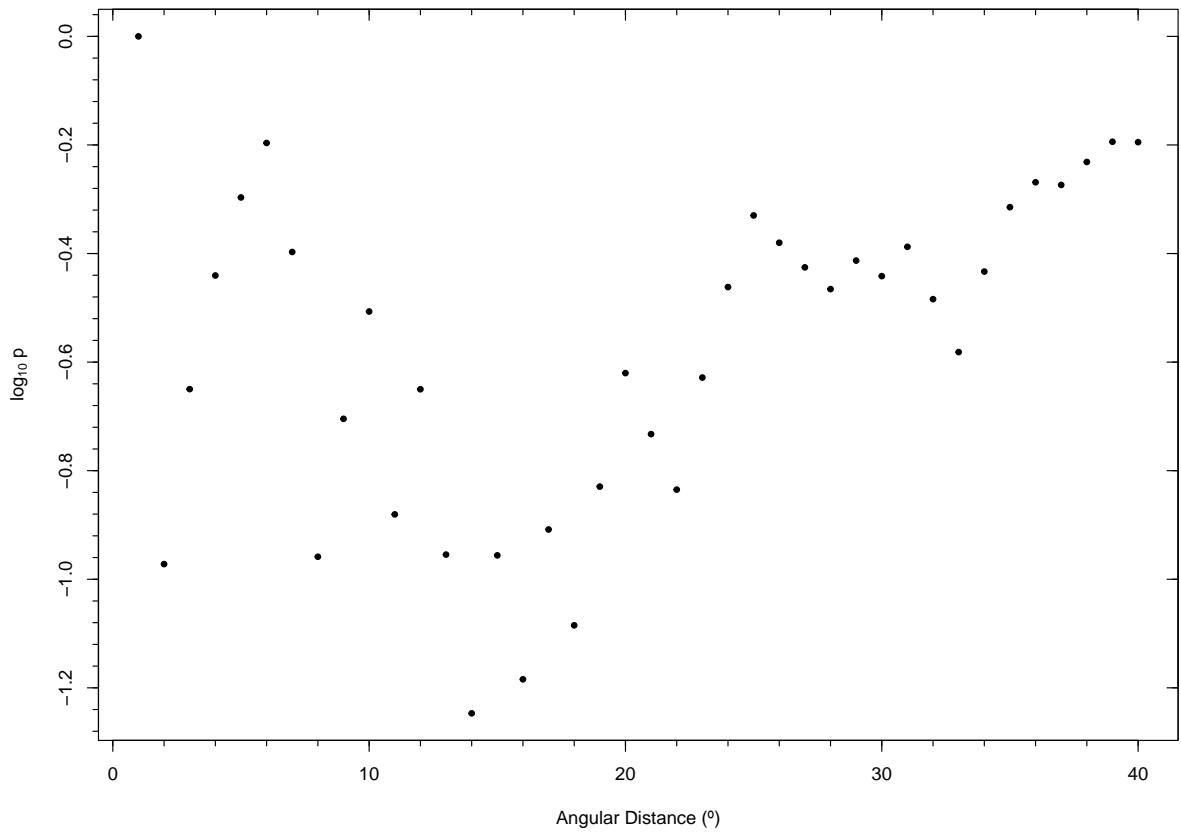}
  \caption{Points represent the ${\rm p}$-values as a function of different angular distances from Cen B.  The Poisson test was used to study the  probability of cosmic ray counts around Cen B. All the p-values are larger than 0.05.}\label{fig:test}
\end{figure}

In order to estimate the number of UHECRs that could be associated to Cen B, we plot in Figure \ref{Skymap_cenB} a sky-map of the UHECR events with $E_{\rm A}\geq5.8\times10^{19}\, {\rm eV}$ collected by PAO over 10 years of operations, from 2004 January 1$^{\rm th}$ up to 2014 March 31 \citep{2015ApJ...804...15A}.   A blow-up of the sky-map of Cen B (black point) with a black circle of 15$^\circ$ around this radio galaxy  is shown in Figure \ref{SkymapZoom}.  UHECRs and high-energy neutrinos are displayed in red crosses and blue triangles, respectively. The blue contour represents the median angular error of the IC35 neutrino event.  The number of  UHECRs in this region that could have been tentatively accelerated in Cen B and after deviated by Galactic and extragalactic magnetic fields would be 10. However, this number should not be taken into account without doing a robust analysis.  In this subsection the statistical and exhaustive analysis is presented.\\
\\
CRs are rare events described by a Poisson distribution.  The difference between two Poisson event samples has been discussed in \cite{2018MNRAS.481.4461F}. The most popular statistical probe to compare two Poisson event samples is the conditional test given by \cite{prz40}, that is based on the binomial distribution.\\
To perform the conditional test, we define a set of solid angles $\Omega(\theta)$ ranging from $\theta=1,2,\ldots 40$\degree~\\ around Cen~B as {\small $\Omega(\theta) = 2 \pi (1 - \cos \theta) \, \mathrm{sr}$}, and a control sample to measure the background that comprises all the cosmic rays detected in the southern sky hemisphere except the region of 40\degree ~ around the target object.   The solid angle corresponding to the control sample is then {\small $\Theta = 2 \pi - \Omega(40\degree) = 4.813199 \, \mathrm{sr} \equiv 15800.8 \deg^2$}.
The lapse of time that a given patch of the sky is monitored depends on both the declination of the patch and the latitude of the observatory. For the Pierre Auger observatory located at $35\degree\, 28'\, 28.05"$ S and $69\degree\, 35'\,7.06"$ W\footnote{https://en.wikipedia.org/wiki/Pierre\_Auger\_Observatory}, the sky circumpolar cap comprises declinations $\delta \lesssim -20\degree$~and it is monitored continuously, while the spherical segment between declinations $\delta \in \pm 20$\degree ~ is visible only a fraction of the day.
As shown in \cite{2018MNRAS.481.4461F}, each cosmic ray event is weighted by a factor $\omega$ given by: $\omega = 1$ if $\delta \leq -90\degree - \mathrm{L}$ and  $\omega = \frac{\pi}{\arccos(-\tan \mathrm{\delta})}$ in other case.\\
Cosmic rays above $\delta > 0$ are excluded to prevent any bias introduced by poor monitored regions.  Thus, all the cosmic rays considered have weights
 $w < 2.$ 
The binomial test procedure is carried on comparing the ray weighted counts $n_\Omega$ in the $\Omega$ solid angle around Cen~B, with the expected number of weighted counts $N$ in the solid angle $\Omega + \Theta$. The null hypothesis is that the number of counts $n_\Omega$ in the $\Omega$ region is distributed like the background. Thus, the probability of a cosmic ray hit is proportional to the solid angle $\Omega$ and given by {\small $p(\Omega) = \frac{\Omega}{\Omega + \Theta}$}.
\noindent Then, the $n_\Omega$ counts are binomial distributed $n_\Omega \sim \mathrm{Binomial}[N,p(\Omega)].$
For the alternative hypothesis we consider the case for which the counts $n_\Omega$ are significantly larger than the expected number $\mathrm{E}(n_\Omega) = N p(\Omega).$ We choose a significance level of $\alpha = 0.001,$ which is the probability of a null hypothesis false rejection or Type I error, and we perform the test using the binomial distribution. This test can be computed as in Fraija et al. \cite{2018MNRAS.481.4461F}.
%
%
\noindent Note the use of the function round to convert the cosmic ray weighted counts to their approximate integer values.
We repeat this procedure for different values of the angular distance to check the possible excess of CRs around Cen~B. Figure \ref{fig:test} shows the logarithm of the binomial test ${\rm p}$-values vs. the angular distance.
None of these probabilities are smaller than the significance level $\alpha = 0.001$ ($\log \alpha = -3$), thus we do not find evidence to reject the null hypothesis, i.e. no excess of CRs is detected around Cen~B.
\subsection{CR Luminosities}
Taking into consideration the null result found from the statistical analysis (see Section \ref{statistical}), we could specify an upper limit based on the non-observation of a clustering near Cen B.\\
To determine the upper limit of the CR luminosity {\small$ L_A= 4\,\pi\,d^2_z A_A\,\int_{E_{\rm A,th}} \,E_A\,E_A^{-\alpha_p}  dE_p\,$},  we use the definition of the expected number of UHECRs $N^{\rm obs}_{\rm cr}$= (PAO exposure) $\times N_{\rm A}$ where  the PAO  exposure  for a point source is given by $\Xi\, \omega(\delta_s)/\Omega_{60}=(6.6\times10^4 \times 0.64/\pi$) km$^2$yr.  Here, $\omega(\delta_s)$ is a correction factor for the declination of Cen B, $E_{\rm A,th}=5.8\times10^{19}\, {\rm eV}$ is the threshold energy, and  $\alpha_A$ and $A_A$ are the spectral power-law index and the proportionality constant, respectively,  of the simple power law assumed  $dN_{\rm A}/dE_A=  A_A\, E^{-\alpha_A}$.  Given the expected number of UHECRs and the PAO  exposure,  the upper limit of the CR luminosity becomes 
{\small
\bary
L_A&=&  4.4\times 10^{39}\,\,{\rm  erg/s}\left(\frac{\alpha_{\rm A} - 1}{\alpha_{A} - 2}\right) \left(\frac{E_{\rm A,th}}{5.8\times10^{19}\, {\rm eV}}\right)^{-1+\alpha_{A}} \left(\frac{N^{\rm obs}_{\rm cr}}{1}\right)\,\cr
&& \hspace{4.8cm}\left(\frac{E_A}{6.0\times10^{19}\, {\rm eV}} \right)^{2-\alpha_{\rm p}}\,.
\label{num}
\eary
}
Table \ref{luminosity} displays {\rm the upper limits} of the UHECR luminosities at 1 EeV (column 4), 60 PeV (column 3) and 6 TeV (column 2) for power indexes  $\alpha_A$= 2.2, 2.4 and 2.6.   It is worth noting that the production of $1.5\times10^{40}$ erg/s in UHECRs by Cen B, one of the dominant radio galaxies within $\sim$ 60 Mpc together Cen A, represents an UHECR emissivity of $8.8\times 10^{41}\,{\rm erg/yr/Mpc^3}$.\\
The upper limit of the CR luminosities at 60 PeV  and 6 TeV are calculated extrapolating down the upper limit  of the UHECR luminosities. We want to emphasize that if  we would have considered one UHECR event (the event within a circular of $3^\circ$ associated by \citep{2009ApJ...693.1261M}) over this radio galaxy, the upper limits reported in Table \ref{luminosity} would change to be the CR luminosities.\\
\section{Neutrinos}
In the region close to the core and within the extended  lobes, accelerated CRs interact  with external radiation fields and ambient gas whereas they propagate along their paths.   Given the CR spectrum, the neutrino spectrum in the case of photo-hadronic and hadronic interactions is  \citep{2008PhR...458..173B, 2014MNRAS.437.2187F}
{\small 
\begin{equation}
\label{Lnu}
E_\nu\, L_\nu  \simeq E_A\,L_A \times \left\lbrace\begin{matrix}
\frac38\, {\rm f_{A\gamma}} &  \hspace{1.3cm}{\rm for\, photo-hadronic \ interactions},\\
\ \ \ {\rm f_{Ap}} & {\rm for\, hadronic\ interactions},
\end{matrix}\right.
\end{equation}
\small}
with
{\small $f_{\rm A\gamma}\simeq {\rm min} \{1.5\times 10^{-5}\, A\,\left( \frac{n_\gamma}{10^6\,\rm cm^{-3}} \right)\left(\frac{\rm r_d}{10^{17}\,{\rm cm} } \right), 1\}$}  \citep{ste68, PhysRevLett.78.2292, 2014PhRvD..90b3007M} and  {\small $f_{Ap}\simeq {\rm min} \{0.14\, A^\frac34\,\left(\frac{n_p}{cm^{-3}} \right) \left( \frac{\tau_{\rm  lobes}}{10\, {\rm Myr}}\right),\,1\}$}  \citep{2002MNRAS.332..215A, 2012ApJ...753...40F}  are the photopion efficiency for photo-hadronic and hadronic interactions, respectively.  The values of the photon density ($n_\gamma$), the size of emitting region close to the core ($r_d$) and the age of the lobes ($\tau_{\rm  lobes}$) are obtained from the fit of the broadband SED detected close to the core and in the direction of the extended  lobes.\\
The number expected of shower-type neutrino events in the IceCube telescope is \citep{2015APh....70...54F, 2015APh....71....1F, 2015MNRAS.450.2784F}
\be\label{Nev}
N_{ev} \approx\,T \,\int_{E_{\rm \nu, min}}^{E_{\rm \nu, max}}\,A_{\rm eff} \, \frac{dN_{\nu}}{d\en}\,dE_\nu \,,
\ee
where $T\simeq$ 6 years corresponds the period reported in the HESE catalog,  $E_{\rm \nu, min}=30\, {\rm TeV}$ is the minimum neutrino energy,   $E_{\rm \nu, max}=10\, {\rm PeV}$ is the maximum neutrino energy,  $A_{\rm eff}$ is the neutrino effective area for the shower-type events  in the declination of Cen B and $dN_\nu/dE_\nu$ is the neutrino spectrum which is related with the CR luminosity (see eq. \ref{Lnu}).\\
\subsection{The region close to the core}
Considering the values reported in Table 2, ultra-relativistic CRs co-accelerated together with electrons inside the emitting region can interact with synchrotron photons at $\sim$ 20 eV and hence might create  the 2-PeV neutrino event. However,  using the photon density $n_\gamma\simeq\frac{d^2_z\,F_{\rm ph}}{r^2_d\, E_{\rm pk}}\simeq8.6\times 10^5\,{\rm cm^{-3}}$  associated to the photon flux  $F_{\rm ph}\simeq 10^{-12}\, {\rm erg\,cm^{-2}\,s^{-1}}$, we find that the value of photopion efficiency  $1.1\times 10^{-5}$ is very low. In this case,  using the CR luminosities reported in Table \ref{luminosity},  the number of PeV-neutrino events as a function of nuclei power index is estimated, as shown in Figure \ref{events}.  The dotted orange line shows the number of events created by the photo-hadronic interactions associated to the synchrotron photons close to the core. This dotted line displays that the number of events is much less than one. 
\subsection{Inside  lobes}
\begin{figure}[hbt]
  \includegraphics[height=0.8\linewidth]{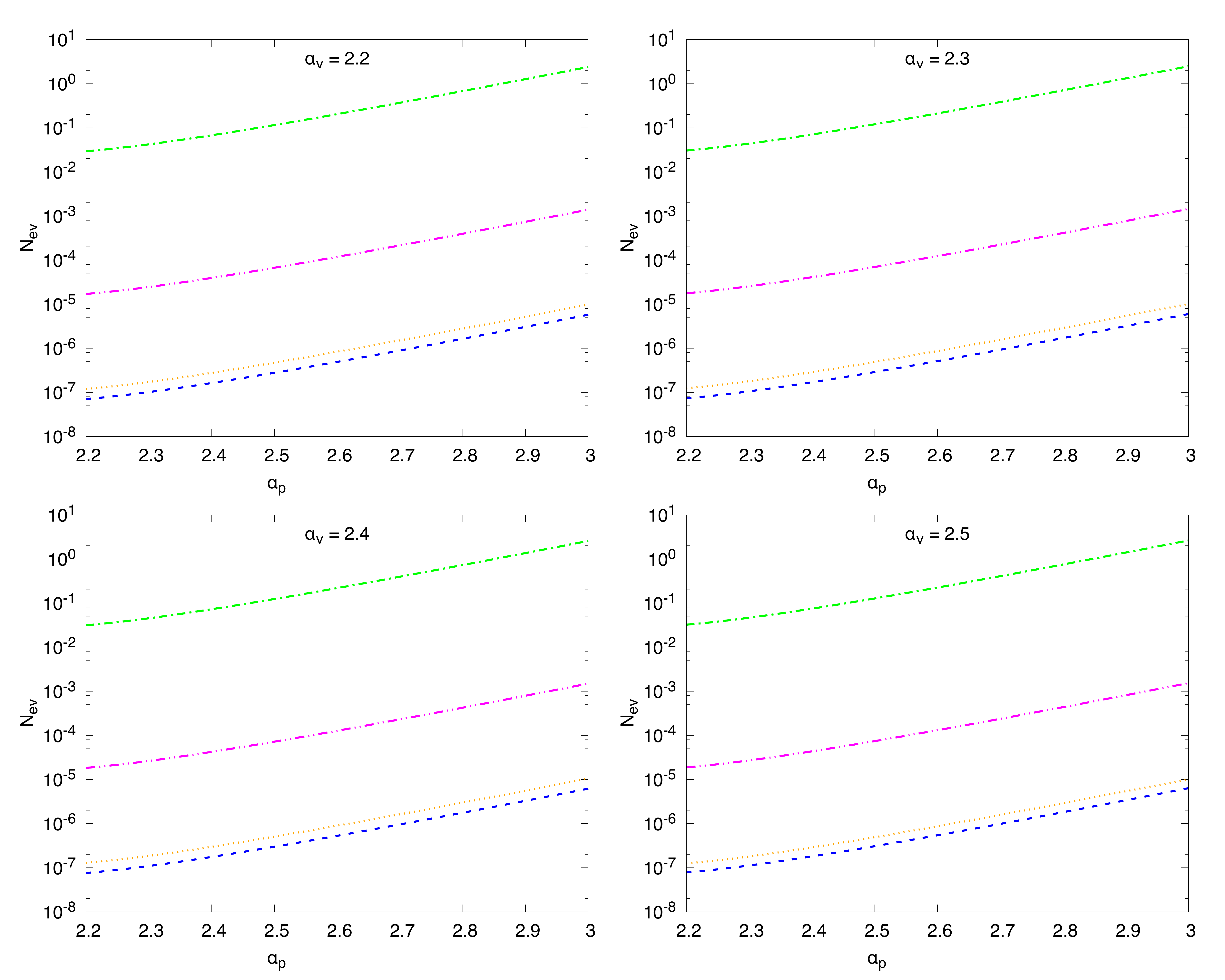}
  \caption{Number of 2-PeV neutrino events as a function of spectral indexes of nuclei (Carbon) and neutrino fluxes.  This number is obtained considering the photo-hadronic interactions associated to synchrotron photons close to the core  (dotted orange line) and  IR photons (dashed blue line),  and the hadronic interactions  associated to a proton density within lobes of $n=10^{-2}\, {\rm cm^{-3}}$ (double dotted-dashed magenta line). The dotted-dashed green line represents the number of events considering an unrealistic/inconceivable density of $n\gtrsim 20\, {\rm cm^{-3}}$.}\label{events}
\end{figure}

\paragraph{Hadronic interactions.}   In accordance with the CR luminosities reported in Table \ref{luminosity} and the CR luminosities normalized with the GeV gamma-ray spectrum through $\pi^0$ decay products (see Fig. \ref{proton_luminosity}),  the average density of target protons is  $n_{\rm p}\simeq10^{-2}\,{\rm cm^{-3}}$.   In this case,  the hadronic efficiency becomes  $5.8\times 10^{-4}$ and the number of PeV-neutrino events as a function of the spectral index of the CR flux is estimated, as shown in Figure \ref{events}.  The double dotted-dashed magenta line in Figure \ref{events} shows the 2-PeV neutrino events as a function of spectral indexes of the CR and neutrino fluxes.  This double dotted-dashed line  exhibits that the number of events lie in the range of $\sim\,\,(10^{-5} - 10^{-3})$. \\
To obtain the 2-PeV neutrino event, an unrealistic/inconceivable density of $20\,{\rm cm^{-3}}$ is required as indicated with the dotted-dashed green line in Figure \ref{events}.\\
\paragraph{Photo-hadronic interactions.} Relativistic CRs within the extended lobes interact with CMB and EBL photon fields.   The photopion efficiencies when the seed photons are those from EBL are  $6.4\times10^{-6}$,  $2.1\times10^{-7}$ and $3.1\times10^{-9}$ for infrared, optical and ultraviolet bands, respectively, and the photopion efficiency when the seed photons are those from CMB  is $1.1\times10^{-2}$.  Using the CR luminosity at 60 PeV reported in Table \ref{luminosity}, the number of PeV-neutrino events as a function of the spectral index of the CR flux is also estimated, as shown in Figure \ref{events}. The dashed blue line shows the number of events created by the photo-hadronic interactions when the seed photons are in the IR band.  This dashed line shows that the number of events lie in the range of $\sim\,\,(10^{-7} - 10^{-5})$.   The number of PeV neutrinos coming from the CR interactions with optical and UV photons is not shown because this is less than $10^{-8}$.
\section{Conclusions}
In this paper, we have analyzed the Fermi-LAT data since August 2008 up to June 2017 around the position of Cen B with coordinates RA=206.7$^\circ$ and Dec 60.41$^\circ$ (J2000).  The latest version v10r0p5 of the publicly available "ScienceTools", the Pass 8 event-level analysis and the instrument response functions have been used in order to obtain the LAT light curve and the gamma-ray spectrum  above 200 MeV.  Cen B has been detected in the LAT energy range with a significance of 19$\sigma$ (TS=365.5).   The best-fit curve of the LAT gamma-ray flux was a power-law function with an integrated flux of $(2.11 \pm 0.19)\times 10^{-8}$ cm$^{-2}$ s$^{-1}$ and a spectral photon index of $2.45\pm 0.06$.\\
Using the broadband SED, we describe the observed gamma-ray flux assuming that this emission is produced close to the core or in the extended lobes. Within the extended  lobes,  the observed gamma-ray flux is modeled with a leptonic and a hadronic scenario. In the leptonic scenario, we require the contributions of  the synchrotron emission and the external IC scattering from the photon fields of the  CMB,  EBL and starlight of the host galaxy. In the hadronic scenario,  we consider the hadronic interactions between the relativistic CRs and the target proton density inside the  lobes.  In the region  close to the core,  the observed gamma-ray flux is fitted using the one-zone SSC model.\\
Based on spectral arguments used to describe the broadband SED of the extended lobes we found that the age of the extended  lobes is $4.1 {\rm Myr}$, the ratio between the magnetic field and the electron density is $4.7$ and the electron density was $1.6\times 10^{-7}\,{\rm cm^{-3}}$ for the values of the magnetic field and the Doppler factor of   $3.8\,{\rm \mu G}$ and $\delta_D=1.5$, respectively.  The value of   the ratio between the magnetic field and electron density suggests that an equipartition mechanism is present in the extended lobes.  The age of the extended  lobes indicates that the  lobes of the radio galaxy Cen A with $\approx 30 {\rm Myr}$ \citep{2009MNRAS.393.1041H, 2014ApJ...783...44F}  are older than the  lobes of Cen B.  In analogy with the two LAT-detected  lobes in radio galaxies Cen A \citep{2010Sci...328..725A} and NGC 6251 \citep{2012ApJ...749...66T},  the description of  the observed gamma-ray flux  by relativistic electrons could imply efficient in-situ acceleration within the extended lobes.\\
After modeling  the broadband SED detected in the region close to the core with the one-zone SSC model,   the best-fit values found are: the Doppler factor $\delta_D=2.2$, the magnetic field $B=0.1\,{\rm G}$, the size of the emitting region $r_d=1.9\times 10^{17}\,{\rm cm}$  and the electron density  $N_e=5.1\times 10^{2}\,{\rm cm^{-3}}$.  These values are similar to those reported by the three LAT-detected radio galaxies Cen A \citep{2010ApJ...719.1433A}, M87 \citep{2009ApJ...707...55A} and NGC1275 \citep{2009ApJ...699...31A}  which favor a classification of this object as FR I.  An estimation of the energy densities  $U_e/U_B\simeq$ 200 indicates that these quantities are not associated with an equipartition mechanism.  Considering the values of the Doppler factor and the size of emitting region, we found the variability time scale  is $\sim$ one month.\\
\\
Taking into account the hadronic contribution required to model the broadband SED of the extended lobes, we study the CR and neutrino fluxes around the position of Cen B.  Given that CRs at UHEs suffer small deviations, we consider the dataset reported by PAO.  We  study the chemical composition of UHECRs through the energy-loss MFP photo-disintegration processes due to CMB photons and the deflecting angle due to Galactic and extragalactic magnetic fields.  We show that only a large fraction of light nuclei such as Carbon, Nitrogen and Oxygen ($6 \leq Z \leq 8$) would be able to produce an excess within a $15^\circ$ angular radius from this radio galaxy. However, as no such excess is observed, we cannot constrain the composition.\\
%
\\
Statistical analysis of the CR distribution around Cen B was performed, finding that this distribution is not different from the background at a level of significance of 5\%. Considering one event which matches perfectly with the position of Cen B  (the event within a circular of $3^\circ$) \citep{2009ApJ...693.1261M},  we normalize  the CR luminosity and extrapolate it  at lower energies.  By computing  the CR luminosity at dozens of PeVs for Oxygen nuclei,  the number of neutrinos expected in the observatory IceCube were obtained through photo-hadronic and hadronic interactions for several radiation fields and materials as targets inside the extended  lobes.   It was shown that though the hadronic interactions are more efficient in producing neutrinos than photo-hadronic ones, in these scenarios the  neutrino fluxes at PeV are too low to be detected in the IceCube experiment. It is worth noting that if CRs in their paths interact with an unrealistic/inconceivable proton density larger than $\gtrsim 20\, {\rm cm^{-3}}$, one PeV neutrino could be created for a spectral index of CRs higher than 2.9. Based on the correlation of UHECRs events through photo-hadronic and hadronic interactions we discard that the IC 35 event could have been created in Cen B. \\
\\
EBL absorption represents the main cause of opacity for high-energy photons.   Franceschini and Rodighiero \cite{2017A&A...603A..34F} estimated the optical depth $\tau_{\gamma\gamma}$ as a function of the photon energy and the redshift. In this analysis can be inferred that  for a source located at z= 0.0129,   the optical depth  becomes $\tau_{\gamma\gamma}$=1 for photons with  energies as high as $\sim$ 20 TeV. In other words, a VHE gamma-ray flux begins to be attenuated ($\tau_{\gamma}>1$) when the photon energy is larger than $\sim$ 20 TeV.  If  the 2-PeV neutrino would have been created around Cen B, it would have been accompanied with the gamma-ray flux at PeV energies.  Because of the EBL absorption,  the PeV photon flux would have been absorbed and hence a cascade emission at TeV would have been created.  The gamma-ray flux is usually estimated from the cascade emissions resulting from the IC processes of the EBL by  $e^\pm$ pairs created by attenuation. For instance,  the  IC energy of the secondary gamma-rays is calculated  as $\epsilon_{\rm \gamma, IC}\simeq (20\,{\rm TeV}/2m_e)^2\times 10^{-2}\,\,{\rm eV}\simeq 3.8\,\, {\rm TeV}$\footnote{ The term $m_e$ corresponds to the electron mass.} with a CR luminosity proportional to $\sim 10^{42}\,{\rm erg/s}$.  Since a gamma-ray flux at few TeVs  has not been detected from this radio galaxy, we discard to Cen B  as  PeV neutrino  emitter.
\acknowledgments
We thank the anonymous referee for a critical reading of this manuscript and valuable suggestions that helped improve the quality and clarity of this paper.   We also thank Antonio Marinelli, Edilberto Aguilar and Dafne Guetta for useful discussions.  NF  acknowledges  financial  support from UNAM-DGAPA-PAPIIT through grant IA102917 and IA102019. MA received funding from the European Union's Horizon 2020 research and innovation program under the Marie Sk\l{}odowska-Curie grant agreement No 690575 and the Universidad de Costa Rica.
%
%

%
\clearpage
\begin{table*}
\begin{center}\renewcommand{\arraystretch}{1.5}\addtolength{\tabcolsep}{4pt}
\caption{Distributions of pressures and energies in  the extended  lobes of Cen B for $\delta_D$=1.5.}\label{values_lobe}
\begin{tabular}{c c c c c c c c}
 \hline
 
\scriptsize{Target thermal density $\,\,({\rm cm^{-3}})$ } & \scriptsize{$n_p$ }  & \scriptsize{$10^{-2}$} & \scriptsize{$10^{-4}\,$}  \\

\hline\hline
 \scriptsize{Non-thermal pressure ($\times 10^{-12}$  dyn cm$^{-2}$ )}  &  \scriptsize{ P$_{nth}$ } & \scriptsize{$5.9$} & \scriptsize{$29.7$}  \\
 \scriptsize{Proton density energy ($\times 10^{-12}$ erg cm$^{-3}$ )}  &  \scriptsize{$U_p $}  & \scriptsize{$0.3$} & \scriptsize{$24.1$}  \\
 \scriptsize{Total energy ($\times 10^{59}$ erg)}   &  \scriptsize{$E_{tot}$} & \scriptsize{$0.9$}   & \scriptsize{$4.6$} \\
\hline
\end{tabular}
\end{center}
\end{table*}
\begin{table*}
\begin{center}\renewcommand{\arraystretch}{1.5}\addtolength{\tabcolsep}{4pt}
\caption{The best-fit and derived values from our leptonic model. The values of distance $d_z=56$ Mpc and an viewing angle estimated in $\theta \approx 20\degree$ were considered.}\label{values_core}
\begin{tabular}{ c c c c c c}
\hline \hline
Quantities   &       Best-fit values    & & Quantities  & Derived  Values   \\
\hline\hline
$\delta_D$ &     2.2  & & $\Gamma$& 6.3\\
$B$ ({\rm G}) &    0.1 & & $\gamma_{\rm e,max}$& $1.5\times10^8$\\
$N_e$ ({\rm cm$^{-3}$})&    $5.1\times 10^2$ & & $U_B$ (erg/cm$^{3}$)  & $3.9\times10^{-4}$  \\
$r_d$ (cm) &    $1.9\times 10^{17}$ & & $U_e$ (erg/cm$^{3}$)  &  $7.3\times10^{-2}$\\
& &  &  $L_e$ (erg/s)  & $1.2\times10^{45}$ \\
\hline
\hline
 \end{tabular}
\end{center}
\begin{center}

\end{center}
\end{table*}
\begin{table}[ht!]\renewcommand{\arraystretch}{1.45}\addtolength{\tabcolsep}{1.5pt}
\centering
\caption{CR luminosity  at different energies.}\label{luminosity}
\begin{tabular}{ l c c c }
 \hline
 $\alpha_A$         &                &   $L_A\,({\rm erg/s})$        &   \\
                            & $6\, {\rm TeV}$  & $60\, {\rm PeV}$& $1\, {\rm EeV}$ \\

 \hline 
 \hline
\scriptsize{2.2}                              & \scriptsize{1.2$\times10^{42}$}  & \scriptsize{1.9$\times10^{41}$}    & \scriptsize{1.1$\times10^{40}$} \\
\scriptsize{2.4}                              & \scriptsize{1.8$\times10^{43}$}  &  \scriptsize{4.5$\times10^{41}$}    &  \scriptsize{1.5$\times10^{40}$} \\
\scriptsize{2.6}                              & \scriptsize{3.4$\times10^{44}$}  &  \scriptsize{1.4$\times10^{42}$}    &  \scriptsize{2.5$\times10^{40}$} \\
\hline
\end{tabular}
\end{table}
\begin{flushleft}
\scriptsize{
}
\end{flushleft}
\clearpage
\appendix
\section{Radiative Processes in the lobes}
\subsection{Synchrotron Radiation}
The non-thermal radio wavelength is interpreted by synchrotron emission of non-thermal electrons $N_e(\gamma_e)=  N_{0,e}\gamma_e^{-\alpha_e} $ for $\gamma_{e,m}<\gamma_e < \gamma_{e,c}$ and $\gamma_{e,c}    \gamma_e^{-(\alpha_e+1)}$  for $\gamma_{e,c} \leq  \gamma_e$,
\noindent where  $N_{0,e}$ is the number of electron density, $\alpha_e$  is the power index of the electron population and  $\gamma_{e,m(c)}$ is the minimum (cooling) electron Lorentz factor. Considering the photons radiated by synchrotron emission, the synchrotron spectral breaks are \citep{2014ApJ...783...44F}
{\small
\begin{eqnarray}\label{synrad}
\epsilon^{\rm syn}_{\gamma,m} &=& 3.4\times 10^{-9}\,\, {\rm eV}\,\,k^2_e\,\lambda^2_{e,B}\,\delta_D\left(\frac{U_B}{10^{-12}\,{\rm erg\,cm^{-3}}} \right)^{\frac52}\, \left(\frac{N_e}{10^{-8} \,{\rm cm^{-3}} }\right)^{-2},\cr
\epsilon^{\rm syn}_{\gamma,c} &=&  5.5\times 10^{-4}\,\, {\rm eV}\,\delta_D\left(\frac{U_B}{10^{-12}\,{\rm erg\,cm^{-3}}} \right)^{-\frac32} \left(\frac{\tau_{\rm syn}}{10\,{\rm Myr}} \right)^{-2},
\end{eqnarray}
}
\noindent  where $\delta_D$ is the Doppler factor, $k_e=\frac{\alpha_e-2}{\alpha_e-1}$, $\lambda_{e,B}=U_e/U_B$ is the equipartition of the energy density defined by $U_e=m_e \int\gamma_e N_e(\gamma_e)d\gamma_e$,  $U_B=B^2/8\pi$ and $\tau_{\rm syn}$ the synchrotron cooling timescale.  The synchrotron spectrum is derived in \cite{2014ApJ...783...44F}.
\subsection{External inverse Compton scattering emission}
Accelerated electrons could scatter external photons up to the range of MeV - GeV energies given by $\epsilon^{\rm ic}_{\rm \gamma, k}\simeq  \gamma^2_{e, c} \epsilon_{\rm \gamma,k}$, where the index ${\rm k}$ represents the CMB, EBL and starlight photon fields. Considering the energies of the photon fields  from CMB, EBL and the starlight of the host galaxy, the inverse Compton spectrum peaks at
{\small
\bary
\label{Aic}
\epsilon^{\rm ic}_{\rm \gamma,CMB} &\simeq& 2.4\times 10^7\,\, {\rm eV}\,\, \left(\frac{U_B}{10^{-12}\,{\rm erg\,cm^{-3}}} \right)^{-2} \left(\frac{\tau_{\rm syn}}{10\,{\rm Myr}} \right)^{-2}\left(\frac{\epsilon_{\rm \gamma,CMB}}{2.5\times 10^{-3}\, {\rm eV} } \right)\,,
\eary
}
{\small
\bary
\label{Aic}
\epsilon^{\rm ic}_{\rm \gamma,EBL} &\simeq& 9.5\times 10^8\,\, {\rm eV}\,\, \left(\frac{U_B}{10^{-12}\,{\rm erg\,cm^{-3}}} \right)^{-2} \left(\frac{\tau_{\rm syn}}{10\,{\rm Myr}} \right)^{-2}\left(\frac{\epsilon_{\rm \gamma,EBL}}{1\, {\rm eV} } \right)\,,
\eary
}
and
{\small
\bary
\label{starlight}
\epsilon^{\rm ic}_{\rm \gamma,star} &\simeq& 5.7\times 10^8\,\, {\rm eV}\,\, \left(\frac{U_B}{10^{-12}\,{\rm erg\,cm^{-3}}} \right)^{-2} \left(\frac{\tau_{\rm syn}}{10\,{\rm Myr}} \right)^{-2}\left(\frac{\epsilon_{\rm \gamma,star}}{0.6\, {\rm eV} } \right)\,,
\eary
}
respectively.
\subsection{Hadronic interactions}
Hadronic interactions within the extended lobes have been proposed to produce gamma-rays in the GeV - TeV energy range \citep{2012ApJ...753...40F}.  Assuming that CRs are described by  a simple power law, $dN/E_A  = A_A E_p^{-\alpha_A}$ with  $A_A$ the proportionality constant and $\alpha_A$ the  spectral index,   the observed gamma-ray spectrum coming from photo-pion decay products is 
\begin{equation}
\label{pp}
\epsilon^{2}_\gamma\, N_\gamma= A^{\rm had}_{\rm \gamma}\, \epsilon_\gamma^{2-\alpha_p},
\end{equation}
where the proportionality constant is normalized through the CR luminosity $L_A$ given by
{\small
\bary
A^{\rm had}_{\gamma}=  1.4\times 10^{-12} \,{\rm \frac{erg}{cm^2\,s}}\,\,     \delta^{-2}_D\, \left(\frac{\rm n_p}{\rm cm^{-3}} \right) \left(\frac{\rm \tau_{\rm lobe}}{10\, \rm Myr} \right)  \left( \frac{E_{\rm A,min}}{\rm GeV}\right)^{\alpha_p-2}\,\left( \frac{L_{\rm A}}{10^{41}\, \rm erg/s} \right)\,.
\label{App}
\eary
}
Here $n_p$ is the thermal particle density, $\tau_{\rm lobe}$  is the age of the lobe, $E_{\rm A,min}$ corresponds to the minimum energy of CRs.
%

\begin{thebibliography}{10}

\bibitem{2009ApJ...693.1261M}
I.~V. {Moskalenko}, L.~{Stawarz}, T.~A. {Porter}, and C.~C. {Cheung}.
\newblock {On the Possible Association of Ultra High Energy Cosmic Rays With
  Nearby Active Galaxies}.
\newblock {\em \apj}, 693:1261--1274, March 2009.

\bibitem{1988ApJ...326...19L}
D.~{Lynden-Bell}, S.~M. {Faber}, D.~{Burstein}, R.~L. {Davies}, A.~{Dressler},
  R.~J. {Terlevich}, and G.~{Wegner}.
\newblock {Spectroscopy and photometry of elliptical galaxies. V - Galaxy
  streaming toward the new supergalactic center}.
\newblock {\em \apj}, 326:19--49, March 1988.

\bibitem{1999PASA...16...53K}
R.~C. {Kraan-Korteweg} and P.~A. {Woudt}.
\newblock {Overview of Uncovered and Suspected Large-scale Structures behind
  the Milky Way}.
\newblock {\em \pasa}, 16:53--59, April 1999.

\bibitem{2000A&ARv..10..211K}
R.~C. {Kraan-Korteweg} and O.~{Lahav}.
\newblock {The Universe behind the Milky Way}.
\newblock {\em \aapr}, 10:211--261, 2000.

\bibitem{2001MNRAS.325..817J}
P.~A. {Jones}, B.~D. {Lloyd}, and W.~B. {McAdam}.
\newblock {The radio galaxy Centaurus B}.
\newblock {\em \mnras}, 325:817--825, August 2001.

\bibitem{1991PASAu...9..255M}
W.~B. {McAdam}.
\newblock {The core jet and relic radiation in the radio source 1343 - 601}.
\newblock {\em Proceedings of the Astronomical Society of Australia}, 9:255,
  1991.

\bibitem{2011A&A...536A...1P}
{Planck Collaboration}, P.~A.~R. {Ade}, N.~{Aghanim}, M.~{Arnaud},
  M.~{Ashdown}, J.~{Aumont}, C.~{Baccigalupi}, M.~{Baker}, A.~{Balbi}, A.~J.
  {Banday}, and et~al.
\newblock {Planck early results. I. The Planck mission}.
\newblock {\em \aap}, 536:A1, December 2011.

\bibitem{2011A&A...536A...7P}
{Planck Collaboration}, P.~A.~R. {Ade}, N.~{Aghanim}, M.~{Arnaud},
  M.~{Ashdown}, J.~{Aumont}, C.~{Baccigalupi}, A.~{Balbi}, A.~J. {Banday},
  R.~B. {Barreiro}, and et~al.
\newblock {Planck early results. VII. The Early Release Compact Source
  Catalogue}.
\newblock {\em \aap}, 536:A7, December 2011.

\bibitem{1998ApJ...499..713T}
M.~{Tashiro}, H.~{Kaneda}, K.~{Makishima}, N.~{Iyomoto}, E.~{Idesawa},
  Y.~{Ishisaki}, T.~{Kotani}, T.~{Takahashi}, and A.~{Yamashita}.
\newblock {Evidence of Energy Nonequipartition between Particles and Fields in
  Lobes of the Radio Galaxy PKS 1343-601 (Centaurus B)}.
\newblock {\em \apj}, 499:713--718, May 1998.

\bibitem{2005ApJS..156...13M}
H.~L. {Marshall}, D.~A. {Schwartz}, J.~E.~J. {Lovell}, D.~W. {Murphy}, D.~M.
  {Worrall}, M.~{Birkinshaw}, J.~M. {Gelbord}, E.~S. {Perlman}, and D.~L.
  {Jauncey}.
\newblock {A Chandra Survey of Quasar Jets: First Results}.
\newblock {\em \apjs}, 156:13--33, January 2005.

\bibitem{2013A&A...550A..66K}
J.~{Katsuta}, Y.~T. {Tanaka}, {\L}.~{Stawarz}, S.~P. {O'Sullivan}, C.~C.
  {Cheung}, J.~{Kataoka}, S.~{Funk}, T.~{Yuasa}, H.~{Odaka}, T.~{Takahashi},
  and J.~{Svoboda}.
\newblock {Fermi-LAT and Suzaku observations of the radio galaxy Centaurus B}.
\newblock {\em \aap}, 550:A66, February 2013.

\bibitem{2009ApJ...697.1071A}
W.~B. {Atwood}, A.~A. {Abdo}, M.~{Ackermann}, W.~{Althouse}, B.~{Anderson},
  M.~{Axelsson}, L.~{Baldini}, J.~{Ballet}, D.~L. {Band}, G.~{Barbiellini}, and
  et~al.
\newblock {The Large Area Telescope on the Fermi Gamma-Ray Space Telescope
  Mission}.
\newblock {\em \apj}, 697:1071--1102, June 2009.

\bibitem{2007Sci...318..938P}
{Pierre Auger Collaboration} and et~al.
\newblock {Correlation of the Highest-Energy Cosmic Rays with Nearby
  Extragalactic Objects}.
\newblock {\em Science}, 318:938--, November 2007.

\bibitem{2008APh....29..188P}
{Pierre Auger Collaboration} and et~al.
\newblock {Correlation of the highest-energy cosmic rays with the positions of
  nearby active galactic nuclei}.
\newblock {\em Astroparticle Physics}, 29:188--204, April 2008.

\bibitem{2015ApJ...804...15A}
A.~{Aab}, P.~{Abreu}, M.~{Aglietta}, E.~J. {Ahn}, I.~A. {Samarai}, I.~F.~M.
  {Albuquerque}, I.~{Allekotte}, J.~{Allen}, P.~{Allison}, A.~{Almela}, and
  et~al.
\newblock {Searches for Anisotropies in the Arrival Directions of the Highest
  Energy Cosmic Rays Detected by the Pierre Auger Observatory}.
\newblock {\em \apj}, 804:15, May 2015.

\bibitem{2017arXiv171001191I}
{IceCube Collaboration}, M.~G. {Aartsen}, M.~{Ackermann}, J.~{Adams}, J.~A.
  {Aguilar}, M.~{Ahlers}, M.~{Ahrens}, I.~A. {Samarai}, D.~{Altmann},
  K.~{Andeen}, and et~al.
\newblock {The IceCube Neutrino Observatory - Contributions to ICRC 2017 Part
  II: Properties of the Atmospheric and Astrophysical Neutrino Flux}.
\newblock {\em ArXiv e-prints}, October 2017.

\bibitem{2011ApJ...743..171A}
M.~{Ackermann}, M.~{Ajello}, A.~{Allafort}, E.~{Antolini}, W.~B. {Atwood},
  M.~{Axelsson}, and {et al.}
\newblock {The Second Catalog of Active Galactic Nuclei Detected by the Fermi
  Large Area Telescope}.
\newblock {\em \apj}, 743:171, December 2011.

\bibitem{2016NatPh..12..807K}
M.~{Kadler}, F.~{Krau{\ss}}, K.~{Mannheim}, R.~{Ojha}, C.~{M{\"u}ller},
  R.~{Schulz}, G.~{Anton}, W.~{Baumgartner}, T.~{Beuchert}, S.~{Buson},
  B.~{Carpenter}, T.~{Eberl}, P.~G. {Edwards}, D.~{Eisenacher Glawion},
  D.~{Els{\"a}sser}, N.~{Gehrels}, C.~{Gr{\"a}fe}, S.~{Gulyaev}, H.~{Hase},
  S.~{Horiuchi}, C.~W. {James}, A.~{Kappes}, A.~{Kappes}, U.~{Katz},
  A.~{Kreikenbohm}, M.~{Kreter}, I.~{Kreykenbohm}, M.~{Langejahn}, K.~{Leiter},
  E.~{Litzinger}, F.~{Longo}, J.~E.~J. {Lovell}, J.~{McEnery}, T.~{Natusch},
  C.~{Phillips}, C.~{Pl{\"o}tz}, J.~{Quick}, E.~{Ros}, F.~W. {Stecker},
  T.~{Steinbring}, J.~{Stevens}, D.~J. {Thompson}, J.~{Tr{\"u}stedt}, A.~K.
  {Tzioumis}, S.~{Weston}, J.~{Wilms}, and J.~A. {Zensus}.
\newblock {Coincidence of a high-fluence blazar outburst with a PeV-energy
  neutrino event}.
\newblock {\em Nature Physics}, 12:807--814, August 2016.

\bibitem{1996ApJ...461..396M}
J.~R. {Mattox}, D.~L. {Bertsch}, J.~{Chiang}, B.~L. {Dingus}, S.~W. {Digel},
  J.~A. {Esposito}, J.~M. {Fierro}, R.~C. {Hartman}, S.~D. {Hunter},
  G.~{Kanbach}, D.~A. {Kniffen}, Y.~C. {Lin}, D.~J. {Macomb}, H.~A.
  {Mayer-Hasselwander}, P.~F. {Michelson}, C.~{von Montigny}, R.~{Mukherjee},
  P.~L. {Nolan}, P.~V. {Ramanamurthy}, E.~{Schneid}, P.~{Sreekumar}, D.~J.
  {Thompson}, and T.~D. {Willis}.
\newblock {The Likelihood Analysis of EGRET Data}.
\newblock {\em \apj}, 461:396, April 1996.

\bibitem{2015ApJS..218...23A}
F.~{Acero}, M.~{Ackermann}, M.~{Ajello}, A.~{Albert}, W.~B. {Atwood},
  M.~{Axelsson}, L.~{Baldini}, J.~{Ballet}, G.~{Barbiellini}, D.~{Bastieri},
  A.~{Belfiore}, R.~{Bellazzini}, E.~{Bissaldi}, and et~al.
\newblock {Fermi Large Area Telescope Third Source Catalog}.
\newblock {\em \apjs}, 218:23, June 2015.

\bibitem{2017arXiv170200664T}
{The Fermi-LAT Collaboration}.
\newblock {3FHL: The Third Catalog of Hard Fermi-LAT Sources}.
\newblock {\em ArXiv e-prints}, February 2017.

\bibitem{2003MNRAS.342.1117M}
T.~{Mauch}, T.~{Murphy}, H.~J. {Buttery}, J.~{Curran}, R.~W. {Hunstead},
  B.~{Piestrzynski}, J.~G. {Robertson}, and E.~M. {Sadler}.
\newblock {SUMSS: a wide-field radio imaging survey of the southern sky - II.
  The source catalogue}.
\newblock {\em \mnras}, 342:1117--1130, July 2003.

\bibitem{2010ApJS..188..405A}
A.~A. {Abdo}, M.~{Ackermann}, M.~{Ajello}, A.~{Allafort}, E.~{Antolini}, W.~B.
  {Atwood}, M.~{Axelsson}, L.~{Baldini}, J.~{Ballet}, G.~{Barbiellini}, and
  et~al.
\newblock {Fermi Large Area Telescope First Source Catalog}.
\newblock {\em \apjs}, 188:405--436, June 2010.

\bibitem{1997NIMPA.389...81B}
R.~{Brun} and F.~{Rademakers}.
\newblock {ROOT - An object oriented data analysis framework}.
\newblock {\em Nuclear Instruments and Methods in Physics Research A},
  389:81--86, February 1997.

\bibitem{2017ApJS..232....7F}
N.~{Fraija}, E.~{Ben{\'{\i}}tez}, D.~{Hiriart}, M.~{Sorcia}, J.~M. {L{\'o}pez},
  R.~{M{\'u}jica}, J.~I. {Cabrera}, J.~A. {de Diego}, M.~{Rojas-Luis}, F.~A.
  {Salazar-V{\'a}zquez}, and A.~{Galv{\'a}n-G{\'a}mez}.
\newblock {Long-term Optical Polarization Variability and Multiwavelength
  Analysis of Blazar Mrk 421}.
\newblock {\em \apjs}, 232:7, September 2017.

\bibitem{2016ApJ...826...31F}
N.~{Fraija} and M.~{Araya}.
\newblock {The Gigaelectronvolt Counterpart of VER J2019+407 in the Northern
  Shell of the Supernova Remnant G78.2+2.1 ({\ensuremath{\gamma}} Cygni)}.
\newblock {\em \apj}, 826(1):31, Jul 2016.

\bibitem{2008A&A...487..837F}
A.~{Franceschini}, G.~{Rodighiero}, and M.~{Vaccari}.
\newblock {Extragalactic optical-infrared background radiation, its time
  evolution and the cosmic photon-photon opacity}.
\newblock {\em \aap}, 487:837--852, September 2008.

\bibitem{2013SAAS...40..225D}
C.~D. {Dermer}.
\newblock {Sources of GeV Photons and the Fermi Results}.
\newblock {\em Astrophysics at Very High Energies, Saas-Fee Advanced Course,
  Volume 40.~ISBN 978-3-642-36133-3.~Springer-Verlag Berlin Heidelberg, 2013,
  p.~225}, 40:225, 2013.

\bibitem{1977A&A....59L...3L}
S.~{Laustsen}, H.-E. {Schuster}, and R.~M. {West}.
\newblock {Probable optical identification of the strong southern radiosource
  13S6A}.
\newblock {\em \aap}, 59:L3, July 1977.

\bibitem{2007A&A...466..481S}
A.~C. {Schr{\"o}der}, G.~A. {Mamon}, R.~C. {Kraan-Korteweg}, and P.~A. {Woudt}.
\newblock {The highly obscured region around <ASTROBJ>PKS 1343 - 601</ASTROBJ>.
  I. Galactic interstellar extinctions using DENIS galaxy colours}.
\newblock {\em \aap}, 466:481--499, May 2007.

\bibitem{2010ApJ...719.1433A}
A.~A. {Abdo} and et~al.
\newblock {Fermi Large Area Telescope View of the Core of the Radio Galaxy
  Centaurus A}.
\newblock {\em \apj}, 719:1433--1444, August 2010.

\bibitem{2014MNRAS.441.1209F}
N.~{Fraija}.
\newblock {Gamma-ray fluxes from the core emission of Centaurus A: a puzzle
  solved}.
\newblock {\em \mnras}, 441:1209--1216, June 2014.

\bibitem{2016ApJ...830...81F}
N.~{Fraija} and A.~{Marinelli}.
\newblock {Neutrino, {$\gamma$}-Ray, and Cosmic-Ray Fluxes from the Core of the
  Closest Radio Galaxies}.
\newblock {\em \apj}, 830:81, October 2016.

\bibitem{2013ApJ...768...54B}
M.~{B{\"o}ttcher}, A.~{Reimer}, K.~{Sweeney}, and A.~{Prakash}.
\newblock {Leptonic and Hadronic Modeling of Fermi-detected Blazars}.
\newblock {\em \apj}, 768:54, May 2013.

\bibitem{2009ApJ...704...38S}
M.~{Sikora}, {\L}.~{Stawarz}, R.~{Moderski}, K.~{Nalewajko}, and G.~M.
  {Madejski}.
\newblock {Constraining Emission Models of Luminous Blazar Sources}.
\newblock {\em \apj}, 704:38--50, October 2009.

\bibitem{2011ApJ...736..131A}
A.~A. {Abdo}, M.~{Ackermann}, M.~{Ajello}, L.~{Baldini}, J.~{Ballet},
  G.~{Barbiellini}, D.~{Bastieri}, K.~{Bechtol}, R.~{Bellazzini}, B.~{Berenji},
  and et~al.
\newblock {Fermi Large Area Telescope Observations of Markarian 421: The
  Missing Piece of its Spectral Energy Distribution}.
\newblock {\em \apj}, 736:131, August 2011.

\bibitem{2014A&A...562A..12P}
M.~{Petropoulou}, E.~{Lefa}, S.~{Dimitrakoudis}, and A.~{Mastichiadis}.
\newblock {One-zone synchrotron self-Compton model for the core emission of
  Centaurus A revisited}.
\newblock {\em \aap}, 562:A12, February 2014.

\bibitem{2017APh....89...14F}
N.~{Fraija}, A.~{Marinelli}, A.~{Galv{\'a}n-G{\'a}mez}, and E.~{Aguilar-Ruiz}.
\newblock {Modeling the spectral energy distribution of the radio galaxy
  IC310}.
\newblock {\em Astroparticle Physics}, 89:14--22, March 2017.

\bibitem{2009NJPh...11f5016D}
C.~D. {Dermer}, S.~{Razzaque}, J.~D. {Finke}, and A.~{Atoyan}.
\newblock {Ultra-high-energy cosmic rays from black hole jets of radio
  galaxies}.
\newblock {\em New Journal of Physics}, 11(6):065016, June 2009.

\bibitem{2018arXiv181010294K}
O.~{Kobzar}, B.~{Hnatyk}, V.~{Marchenko}, and O.~{Sushchov}.
\newblock {Search for ultra high energy cosmic rays sources. Radiogalaxy Virgo
  A}.
\newblock {\em ArXiv e-prints}, October 2018.

\bibitem{2016MNRAS.455..838K}
B.~{Khiali} and E.~M. {de Gouveia Dal Pino}.
\newblock {High-energy neutrino emission from the core of low luminosity AGNs
  triggered by magnetic reconnection acceleration}.
\newblock {\em \mnras}, 455:838--845, January 2016.

\bibitem{1987ApJ...315..504L}
R.~V.~E. {Lovelace}, J.~C.~L. {Wang}, and M.~E. {Sulkanen}.
\newblock {Self-collimated electromagnetic jets from magnetized accretion
  disks}.
\newblock {\em \apj}, 315:504--535, April 1987.

\bibitem{2013MNRAS.430.2828L}
R.~V.~E. {Lovelace} and P.~P. {Kronberg}.
\newblock {Transmission line analogy for relativistic Poynting-flux jets}.
\newblock {\em \mnras}, 430:2828--2835, April 2013.

\bibitem{2013A&A...558A..19W}
S.~{Wykes}, J.~H. {Croston}, M.~J. {Hardcastle}, J.~A. {Eilek}, P.~L.
  {Biermann}, A.~{Achterberg}, J.~D. {Bray}, A.~{Lazarian}, M.~{Haverkorn},
  R.~J. {Protheroe}, and O.~{Bromberg}.
\newblock {Mass entrainment and turbulence-driven acceleration of ultra-high
  energy cosmic rays in Centaurus A}.
\newblock {\em \aap}, 558:A19, October 2013.

\bibitem{2005Ap&SS.298..115L}
R.~V.~E. {Lovelace}, P.~R. {Gandhi}, and M.~M. {Romanova}.
\newblock {Relativistic Jets from Accretion Disks}.
\newblock {\em \apss}, 298:115--120, July 2005.

\bibitem{1976Natur.262..649L}
R.~V.~E. {Lovelace}.
\newblock {Dynamo model of double radio sources}.
\newblock {\em \nat}, 262:649--652, August 1976.

\bibitem{2006A&A...455..773V}
M.-P. {V{\'e}ron-Cetty} and P.~{V{\'e}ron}.
\newblock {A catalogue of quasars and active nuclei: 12th edition}.
\newblock {\em \aap}, 455:773--777, August 2006.

\bibitem{2014PhRvD..90l2006A}
A.~{Aab}, P.~{Abreu}, M.~{Aglietta}, E.~J. {Ahn}, I.~{Al Samarai}, I.~F.~M.
  {Albuquerque}, I.~{Allekotte}, J.~{Allen}, P.~{Allison}, A.~{Almela}, and
  et~al.
\newblock {Depth of maximum of air-shower profiles at the Pierre Auger
  Observatory. II. Composition implications}.
\newblock {\em \prd}, 90(12):122006, December 2014.

\bibitem{2018MNRAS.481.4461F}
N.~{Fraija}, E.~{Aguilar-Ruiz}, A.~{Galv{\'a}n-G{\'a}mez}, A.~{Marinelli}, and
  J.~A. {de Diego}.
\newblock {Study of the PeV neutrino, {$\gamma$}-rays, and UHECRs around the
  lobes of Centaurus A}.
\newblock {\em \mnras}, 481:4461--4471, December 2018.

\bibitem{2008JPhCS.120f2006D}
C.~D. {Dermer}.
\newblock {The obscured universe}.
\newblock In {\em Journal of Physics Conference Series}, volume 120 of {\em
  Journal of Physics Conference Series}, page 062006, July 2008.

\bibitem{2007arXiv0711.2804D}
C.~D. {Dermer}.
\newblock {On Gamma Ray Burst and Blazar AGN Origins of the Ultra-High Energy
  Cosmic Rays in Light of First Results from Auger}.
\newblock {\em ArXiv e-prints}, November 2007.

\bibitem{2009herb.book.....D}
C.~D. {Dermer} and G.~{Menon}.
\newblock {\em {High Energy Radiation from Black Holes: Gamma Rays, Cosmic
  Rays, and Neutrinos}}.
\newblock 2009.

\bibitem{1999MNRAS.306..371H}
J.~L. {Han}, R.~N. {Manchester}, and G.~J. {Qiao}.
\newblock {Pulsar rotation measures and the magnetic structure of our Galaxy}.
\newblock {\em \mnras}, 306:371--380, June 1999.

\bibitem{1999JHEP...08..022H}
D.~{Harari}, S.~{Mollerach}, and E.~{Roulet}.
\newblock {The toes of the ultra high energy cosmic ray spectrum}.
\newblock {\em Journal of High Energy Physics}, 8:022, August 1999.

\bibitem{2002APh....18..165T}
P.~G. {Tinyakov} and I.~I. {Tkachev}.
\newblock {Tracing protons through the Galactic magnetic field: a clue for
  charge composition of ultra-high-energy cosmic rays}.
\newblock {\em Astroparticle Physics}, 18:165--172, October 2002.

\bibitem{2008NuPhS.175...62H}
J.~L. {Han}.
\newblock {New knowledge of the Galactic magnetic fields}.
\newblock {\em Nuclear Physics B Proceedings Supplements}, 175:62--69, January
  2008.

\bibitem{2003astro.ph..1598C}
J.~M. {Cordes} and T.~J.~W. {Lazio}.
\newblock {NE2001. II. Using Radio Propagation Data to Construct a Model for
  the Galactic Distribution of Free Electrons}.
\newblock {\em ArXiv Astrophysics e-prints}, January 2003.

\bibitem{2002astro.ph..7156C}
J.~M. {Cordes} and T.~J.~W. {Lazio}.
\newblock {NE2001.I. A New Model for the Galactic Distribution of Free
  Electrons and its Fluctuations}.
\newblock {\em ArXiv Astrophysics e-prints}, July 2002.

\bibitem{2007APh....26..378K}
M.~{Kachelrie{\ss}}, P.~D. {Serpico}, and M.~{Teshima}.
\newblock {The Galactic magnetic field as spectrograph for ultra-high energy
  cosmic rays}.
\newblock {\em Astroparticle Physics}, 26:378--386, January 2007.

\bibitem{2008ApJ...681.1279T}
H.~{Takami} and K.~{Sato}.
\newblock {Distortion of Ultra-High-Energy Sky by Galactic Magnetic Field}.
\newblock {\em \apj}, 681:1279--1286, July 2008.

\bibitem{2008PhyS...78d5901F}
D.~{Fargion}.
\newblock {Light nuclei solving the AUGER puzzles: the Cen-A imprint}.
\newblock {\em \physscr}, 78(4):045901, October 2008.

\bibitem{1998A&A...335...19R}
D.~{Ryu}, H.~{Kang}, and P.~L. {Biermann}.
\newblock {Cosmic magnetic fields in large scale filaments and sheets}.
\newblock {\em \aap}, 335:19--25, July 1998.

\bibitem{1999ApJ...514L..79B}
P.~{Blasi}, S.~{Burles}, and A.~V. {Olinto}.
\newblock {Cosmological Magnetic Field Limits in an Inhomogeneous Universe}.
\newblock {\em \apjl}, 514:L79--L82, April 1999.

\bibitem{2009PhRvD..80l3012N}
A.~{Neronov} and D.~V. {Semikoz}.
\newblock {Sensitivity of {$\gamma$}-ray telescopes for detection of magnetic
  fields in the intergalactic medium}.
\newblock {\em \prd}, 80(12):123012, December 2009.

\bibitem{1984ARA&A..22..425H}
A.~M. {Hillas}.
\newblock {The Origin of Ultra-High-Energy Cosmic Rays}.
\newblock {\em \araa}, 22:425--444, 1984.

\bibitem{2014MNRAS.439.2050R}
B.~{Reville} and A.~R. {Bell}.
\newblock {On the maximum energy of shock-accelerated cosmic rays at
  ultra-relativistic shocks}.
\newblock {\em \mnras}, 439:2050--2059, April 2014.

\bibitem{2010MNRAS.402..321L}
M.~{Lemoine} and G.~{Pelletier}.
\newblock {On electromagnetic instabilities at ultra-relativistic shock waves}.
\newblock {\em \mnras}, 402:321--334, February 2010.

\bibitem{prz40}
J.~{Przyborowski} and H.~{Wilenski}.
\newblock {}.
\newblock {\em Biometrika}, 1940.

\bibitem{2008PhR...458..173B}
J.~K. {Becker}.
\newblock {High-energy neutrinos in the context of multimessenger
  astrophysics}.
\newblock {\em \physrep}, 458:173--246, March 2008.

\bibitem{2014MNRAS.437.2187F}
N.~{Fraija}.
\newblock {GeV-PeV neutrino production and oscillation in hidden jets from
  gamma-ray bursts}.
\newblock {\em \mnras}, 437:2187--2200, January 2014.

\bibitem{ste68}
F.~W. {Stecker}.
\newblock {Effect of Photomeson Production by the Universal Radiation Field on
  High-Energy Cosmic Rays}.
\newblock {\em Physical Review Letters}, 21:1016--1018, September 1968.

\bibitem{PhysRevLett.78.2292}
Eli Waxman and John Bahcall.
\newblock High energy neutrinos from cosmological gamma-ray burst fireballs.
\newblock {\em Phys. Rev. Lett.}, 78:2292--2295, Mar 1997.

\bibitem{2014PhRvD..90b3007M}
K.~{Murase}, Y.~{Inoue}, and C.~D. {Dermer}.
\newblock {Diffuse neutrino intensity from the inner jets of active galactic
  nuclei: Impacts of external photon fields and the blazar sequence}.
\newblock {\em \prd}, 90(2):023007, July 2014.

\bibitem{2002MNRAS.332..215A}
F.~A. {Aharonian}.
\newblock {Proton-synchrotron radiation of large-scale jets in active galactic
  nuclei}.
\newblock {\em \mnras}, 332:215--230, May 2002.

\bibitem{2012ApJ...753...40F}
N.~{Fraija}, M.~M. {Gonz{\'a}lez}, M.~{Perez}, and A.~{Marinelli}.
\newblock {How Many Ultra-high Energy Cosmic Rays Could we Expect from
  Centaurus A?}
\newblock {\em \apj}, 753:40, July 2012.

\bibitem{2015APh....70...54F}
N.~{Fraija} and A.~{Marinelli}.
\newblock {TeV {\ensuremath{\gamma}}-ray fluxes from the long campaigns on Mrk
  421 as constraints on the emission of TeV-PeV neutrinos and UHECRs}.
\newblock {\em Astroparticle Physics}, 70:54--61, Oct 2015.

\bibitem{2015APh....71....1F}
N.~{Fraija}.
\newblock {Could a plasma in quasi-thermal equilibrium be associated to the
  ``orphan'' TeV flares?}
\newblock {\em Astroparticle Physics}, 71:1--20, Dec 2015.

\bibitem{2015MNRAS.450.2784F}
Nissim {Fraija}.
\newblock {Resonant oscillations of GeV-TeV neutrinos in internal shocks from
  gamma-ray burst jets inside stars}.
\newblock {\em \mnras}, 450(3):2784--2798, Jul 2015.

\bibitem{2009MNRAS.393.1041H}
M.~J. {Hardcastle}, C.~C. {Cheung}, I.~J. {Feain}, and {\L}.~{Stawarz}.
\newblock {High-energy particle acceleration and production of
  ultra-high-energy cosmic rays in the giant lobes of Centaurus A}.
\newblock {\em \mnras}, 393:1041--1053, March 2009.

\bibitem{2014ApJ...783...44F}
N.~{Fraija}.
\newblock {Correlation of {$\gamma$}-Ray and High-energy Cosmic Ray Fluxes from
  the Giant Lobes of Centaurus A}.
\newblock {\em \apj}, 783:44, March 2014.

\bibitem{2010Sci...328..725A}
A.~A. {Abdo} and et~al.
\newblock {Fermi Gamma-Ray Imaging of a Radio Galaxy}.
\newblock {\em Science}, 328:725--, May 2010.

\bibitem{2012ApJ...749...66T}
Y.~{Takeuchi}, J.~{Kataoka}, {\L}.~{Stawarz}, Y.~{Takahashi}, K.~{Maeda},
  T.~{Nakamori}, C.~C. {Cheung}, A.~{Celotti}, Y.~{Tanaka}, and T.~{Takahashi}.
\newblock {Suzaku X-Ray Imaging of the Extended Lobe in the Giant Radio Galaxy
  NGC 6251 Associated with the Fermi-LAT Source 2FGL J1629.4+8236}.
\newblock {\em \apj}, 749:66, April 2012.

\bibitem{2009ApJ...707...55A}
A.~A. {Abdo} and et~al.
\newblock {Fermi Large Area Telescope Gamma-Ray Detection of the Radio Galaxy
  M87}.
\newblock {\em \apj}, 707:55--60, December 2009.

\bibitem{2009ApJ...699...31A}
A.~A. {Abdo} and et~al.
\newblock {Fermi Discovery of Gamma-ray Emission from NGC 1275}.
\newblock {\em \apj}, 699:31--39, July 2009.

\bibitem{2017A&A...603A..34F}
A.~{Franceschini} and G.~{Rodighiero}.
\newblock {The extragalactic background light revisited and the cosmic
  photon-photon opacity}.
\newblock {\em \aap}, 603:A34, July 2017.

\bibitem{2004AJ....128.2593F}
A.~L. {Fey}, R.~{Ojha}, J.~E. {Reynolds}, S.~P. {Ellingsen}, P.~M. {McCulloch},
  D.~L. {Jauncey}, and K.~J. {Johnston}.
\newblock {Astrometry of 25 Southern Hemisphere Radio Sources from a VLBI
  Short-Baseline Survey}.
\newblock {\em \aj}, 128:2593--2598, November 2004.

\bibitem{2009MNRAS.395..504B}
S.~{Burke-Spolaor}, R.~D. {Ekers}, M.~{Massardi}, T.~{Murphy}, B.~{Partridge},
  R.~{Ricci}, and E.~M. {Sadler}.
\newblock {Wide-field imaging and polarimetry for the biggest and brightest in
  the 20-GHz southern sky}.
\newblock {\em \mnras}, 395:504--517, May 2009.

\end{thebibliography}
%

%
\clearpage
%
%
%
%

\clearpage

\clearpage

\end{document}